\documentclass[aip,chaos,preprint, superscriptaddress,floatfix,tightenlines,showpacs,twocolumn,10pt,notitlepage]{revtex4-2}
\usepackage{amsmath}
\usepackage{blindtext}
\usepackage{amssymb}
\usepackage{amsthm}
\usepackage{bm}
\usepackage[shortlabels]{enumitem}
\usepackage{tabularx}
\usepackage{tabularray}
\usepackage{graphicx}
\usepackage{tikz-cd}
\usepackage{xcolor}
\usepackage{hyperref}
\usepackage[normalem]{ulem}
\usepackage[textsize=scriptsize,backgroundcolor=white]{todonotes}
\usepackage{import}
\renewcommand{\Re}{\textrm{Re}}

\hypersetup{
	colorlinks=true,
	linkcolor=blue,
	citecolor=blue,			%Crimson
	urlcolor=blue,			%Crimson
	filecolor=blue,
	anchorcolor = green,
}

\newcommand{\mb}[1]{\bm{#1}}

\begin{document}
\title{\large{Prediction performance of random reservoirs with different topology for nonlinear dynamical systems with different number of degrees of freedom}}

\author{Shailendra K. Rathor}
\affiliation{Institute of Thermodynamics and Fluid Mechanics, Technische Universit\"at
Ilmenau, P.O.Box 100565, D-98684 Ilmenau, Germany}

\author{Lina Jaurigue}
\affiliation{Institute of Physics, Technische Universit\"at Ilmenau, P.O.Box 100565, D-98684 Ilmenau, Germany}

\author{Martin Ziegler}
\affiliation{Energy Materials and Devices, Department of Materials Science, Faculty of Engineering, Christian-Albrechts-Universität zu Kiel, D-24143 Kiel, Germany}

\author{Jörg Schumacher}
\affiliation{Institute of Thermodynamics and Fluid Mechanics, Technische Universit\"at
Ilmenau, P.O.Box 100565, D-98684 Ilmenau, Germany}

\date{\today}

\begin{abstract}
Reservoir computing (RC) is a powerful framework for predicting nonlinear dynamical systems, yet the role of reservoir topology—particularly symmetry in connectivity and weights—remains not adequately understood. This work investigates how the structure of the network influences the performance of RC in four systems of increasing complexity: the Mackey-Glass system with delayed-feedback, two low-dimensional thermal convection models, and a three-dimensional shear flow model exhibiting transition to turbulence. Using five reservoir topologies in which connectivity patterns and edge weights are controlled independently, we evaluate both direct- and cross-prediction tasks. The results show that symmetric reservoir networks substantially improve prediction accuracy for the convection-based systems, especially when the input dimension is smaller than the number of degrees of freedom. In contrast, the shear-flow model displays almost no sensitivity to topological symmetry due to its strongly chaotic high-dimensional dynamics. These findings reveal how structural properties of reservoir networks affect their ability to learn complex dynamics and provide guidance for designing more effective RC architectures. 
\end{abstract}

\maketitle

\begin{quotation}
{Predicting the evolution of nonlinear dynamical systems remains a central challenge across science and engineering. This study shows that the symmetry of a reservoir computer’s network topology—defined by both its node connections and connection weights—has a decisive impact on prediction accuracy. When the input dimension is smaller than the number of degrees of freedom of the target system, symmetric reservoir topologies yield superior performance due to enhanced cross prediction between variables. In contrast, asymmetric reservoirs perform better only when provided with full information from all degrees of freedom. These results demonstrate how the structure of reservoir networks fundamentally shapes the predictive capabilities of the model and offers new pathways for designing more effective RC architectures.}
\end{quotation}

%%% OUTLINE

\section{Introduction}
Reservoir computing (RC) has become a promising brain-inspired paradigm for recurrent data processing among machine-learning and neural-network architectures.\cite{Jaeger2001,Maass2002,Pathak2018,Lukosevicius2009} Its inherent recurrence generates short-term memory and enables the prediction of complex temporal evolutions, allowing RCs to serve as surrogate models for nonlinear dynamical processes.\cite{Tanaka2019} RC has been applied widely across engineering and the natural sciences,\cite{Zolfagharinejad_2024, Schuman_2022,Coulombe_2017} including neuroscience, complex systems, machine learning, and fluid dynamics.\cite{damicelli2022,Pathak2018,Vlachas2020,Pandey2020,Heyder2021,Doan2021,Heyder2022} Compared with deep recurrent neural networks, RC can be computationally cheaper while maintaining high prediction accuracy.\cite{Yildiz2012,Pandey2020a} A wide range of physical reservoir realizations have been explored, including water-bucket systems,\cite{Tanaka2019} Ising-spin networks,\cite{Cindrak2024} and parametric quantum circuits.\cite{Pfeffer2022} As a result, RC remains an active research area, particularly with respect to physical implementations.\cite{Stepney_2024, Tanaka2019, Huelser2022}

Despite this progress, key structural questions remain open. In particular, the size of the reservoir is crucial: it must be large enough to capture the complexity of the target system while avoiding unnecessary computational cost. This raises the question of what the minimal reservoir size is for a given problem. Additionally, the topology and connectivity of the reservoir strongly influence performance, yet remain poorly understood.\cite{Dale2021} Existing studies investigate various structural aspects, including directionality, sparsity, average path length, clustering, and degree distributions.\cite{Dale2021,Dale2019,Lukosevicius2009} Reservoirs with ring, lattice, random, small-world,\cite{Watts1998,Kawai2019,Rodan_2011} scale-free,\cite{Deng2007} and modular structures\cite{Rodriguez2019} have all been explored. Notably, some studies report that for small reservoirs, uncoupled nodes may enhance long-term prediction and chaotic-attractor reconstruction.\cite{Jaurigue2024a,Ma2023,Yadav_2025,Griffith2019} Recent results also show that random connectivity can outperform structured topologies, such as small-world networks when predicting Mackey–Glass time series.\cite{Rathor2025} Overall, however, the relationship between reservoir topology and performance remains unresolved, and no general principle yet links network structure to predictive capability.\cite{Geier2025,Bingoel2025}

%--------------------------------------------------------
\begin{figure*}
	\includegraphics[scale=0.95]{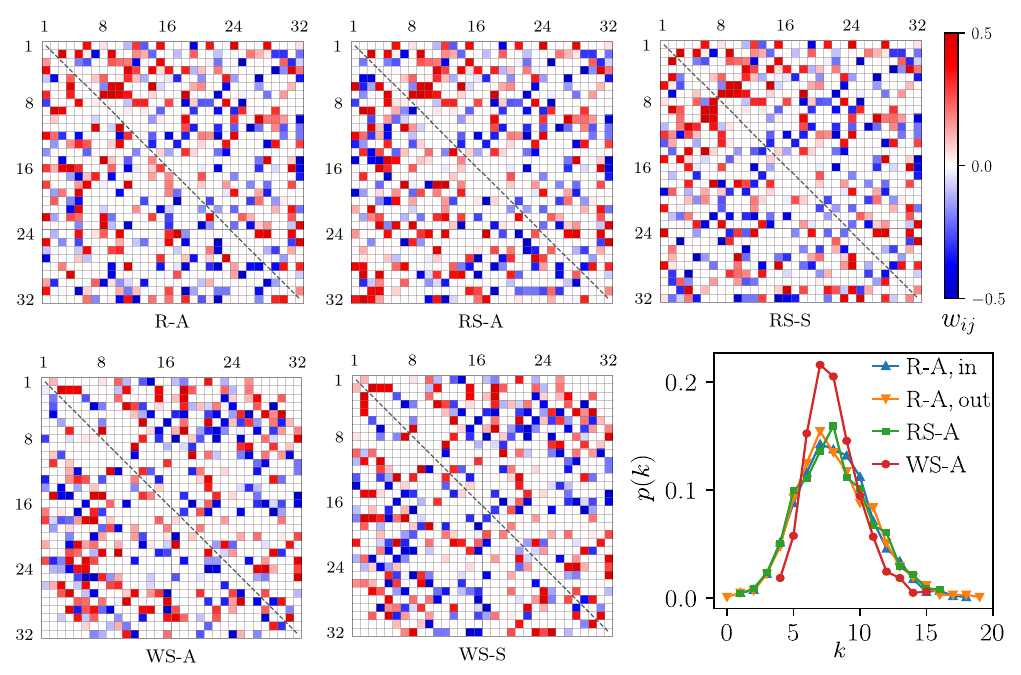}
	\caption{Reservoir matrices $W$ for five different random network topologies, namely, random--asymmetric (R-A), random symmetrized--asymmetric (RS-A), random symmetrized--symmetric (RS-S), Watts-Strogatz--asymmetric (WS-A), and Watts-Strogatz--symmetric (WS-S). We display examples with $N=32$ reservoir nodes for a better visibility. In the subsequent analysis of the dynamical systems $N=1024$ in most cases. A filled square at position $(i,j)$ stands for an active connection from node $i$ to node $j$, i.e., $i\to j$, one at $(j,i)$ for $j\to i$. The color of the square stands for the magnitude of the weight, i.e., $w_{ij}\in \mathbb{R}$ for $i\to j$ and $w_{ji}\in \mathbb{R}$ for $j\to i$. The diagonal is indicated in all plots for better visual guidance with respect to symmetry. Bottom-right: node-degree distribution of network topologies R-A, RS-A, and WS-A with 1024 nodes each. The distributions of incoming and outgoing degrees (in, out) are different for asymmetric connection network R-A while they are same for remaining symmetric connection network topologies. The networks RS-S (WS-S) and RS-A (WS-A) are not shown as they have same node-degree distribution owing to shared connection matrices.}
	\label{fig:network_diagram}
\end{figure*}
%--------------------------------------------------------
%
In this work, we investigate the structure of reservoir network for predicting the dynamics of four nonlinear dynamical systems; three of them are derived as low-dimensional Galerkin models from fluid dynamical systems, such as buoyancy- and shear-driven fluid flows. The first two systems stand for truncations of thermal convection processes in a two-dimensional fluid layer, the Lorenz 63 model and its 8-dimensional extension.\cite{Lorenz1963,Gluhovsky2002} The third is a time-dependent three-dimensional shear flow, which transitions from the laminar state to turbulence.\cite{Moehlis_2004} System No. 4 is a Mackey-Glass equation.\cite{Mackey_1977} We compare their performance in a dynamics prediction task for five different topologies of the reservoir network in which network connectivity and weights along the network edges are chosen separately, similar to Ref.~\onlinecite{Rathor2025}. We find that the dynamics prediction, which comprises of direct- and cross-prediction tasks, is improved if reservoir network is symmetric for the first two systems. However, dynamics prediction of the third system is insensitive to the symmetry of the topology owing to the chaotic dynamics in higher-dimensional space.

The outline of the manuscript is as follows. Section II discusses the reservoir computing model approach and presents the different network topologies. In Section III, we present the four dynamical systems of increasing complexity that we study here. In Sec. IV, we summarize their performance in predicting the dynamics with partial state information provided to the reservoir. This includes a comparison of the five different topologies for the different dynamical systems. We summarize our analysis and give an outlook in Sec. VI.     

\section{Reservoir Network Topologies}
In the reservoir computing model a target output $\mb{y}(t)$ is predicted using its linear relationship with the reservoir state $\mb{r}(t)$ that is driven by the input $\mb{u}(t)$. The reservoir state $\mb{r}(t)$ is a vector representing the state of $N$ neurons and evolves in discrete time steps $\Delta t$, i.e., $t = k\Delta t$ (see \ref{subsec:rc_time_step}), as follows:
 \begin{align}
 	\mb{r}(k+1) &= (1-\varepsilon)\mb{r}(k) + \nonumber \\
                &\varepsilon\tanh \left[W  \mb{r}(k) + W^{\rm in} \mb{u}(k+1)\right]\,,
 	\label{eq:RC}
 \end{align}
where $W^{\rm in}$ is a fixed random matrix to map an $N_{\rm in}$-dimensional input vector ${\bm u}(k+1)$ to the reservoir, $W$ is the reservoir coupling matrix, and $\varepsilon$ is the leaking rate. The state evolves without any bias in the nonlinear term. The predicted output $\mb{y}^p$ is given by the linear relationship,
 \begin{equation}
 	\mb{y}^p(k+1)  = W^{\ast {\rm out}}\mb{r}(k+1).%\nonumber
 	\label{eq:output_layer}
 \end{equation}
The matrix $W^{\ast {\rm out}}$ is the optimized output matrix. Only this matrix is trained with respect to the target output $\mb{y}(k+1)$ using the ridge regression scheme with a Tikhonov regularization parameter of $\gamma$. The elements of the matrix $W^{\rm in}$ are randomly chosen from a uniform distribution ${\cal U}([-0.5,0.5])$. The reservoir state is evolved for the first $t_0$ time steps during which it depends on the initial reservoir state (known as wash-out).~\cite{Jaeger2001,Yildiz2012} The states $\mb{r}(k<k_0)$ are discarded as transients, which depend on the spectral radius of $W$, the leakage rate and the prediction task. The transience $k_0$ depends on reservoir type and hence we take sufficiently large $k_0 = 500$ for all reservoir types and tasks. The states $\mb{r}(k_0<k<k_t)$ are collected and used to train the RC model output layer with matrix $W^{\rm out}$ such that $\mb{y}^p(k)=W^{\rm out}\mb{r}(k)$. The output layer is further optimized for the hyperparameters (spectral radius, leaking rate, and regression parameter) to obtain $W^{\ast {\rm out}}$ in Eq.~(\ref{eq:output_layer}). The optimized output matrix is employed to test the prediction for $k_t< k<k_p$ with testing phase of $k_p-k_t$ time steps. The matrix $W$ is sparse representing a small fixed density of active nodes in all network types in this study.

We define a dynamics prediction task (DPT) as the prediction of the complete system state $\mb{y}(k) = [y_1, y_2, ..., y_{M}]^T \in \mathbb{R}^M$ of the target dynamical system at the next time instant $(k+1)\Delta t$. In other words, a DPT consists of $M$ tasks to predict $M$ time series of the components of $\mb{y}(k)$ in successive one-step predictions. The RC model can be trained using either (i) all $M$ time series $\mb{u} = [u_1, u_2, ..., u_M]^T \in \mathbb{R}^M$ or (ii) $m<M$ time series $\mb{u} = [u_1, u_2, ..., u_{m}]^T \in \mathbb{R}^m$ as the input to predict the next full system state, including the unseen data of remaining dimensions $M-m$. This is also known as the {open-loop scenario} of reservoir computing since the ground truth input is fed at each step.\cite{Lukovsevivcius2012} In the case of (i), the prediction is expected to be easier, since the complete system state information is given to the RC model. In the case of (ii), the RC model requires enough memory for a delay embedding to reconstruct the full state.~\cite{Marquez_2019,Jaurigue_2025,Jaurigue2024,Fleddermann_2025} 

We will focus to RC tasks with partial information as input, investigate the performance and its relation with the reservoir topology. The DPTs of multi-dimensional states, namely Lorenz model (L63), extended Lorenz model (L8) and Galerkin shear flow model (SF), proceed by feeding only 1, 2, and 5 inputs ${\bm u}$, respectively, to predict states ${\bm y}^p$ of dimensions 3, 8, and 9, respectively, including the ones that are not seen by the RC, see Table~\ref{tab:systems_summary}. Finally, we denote a task as {\em direct prediction} if component $u_d$ predicts component $y^p_d$ and as {\em cross prediction} if $u_d$ predicts $y_{d'}$, i.e., when component indices are different, $d\ne d'$. This implies, that open-loop RC tasks with $m<M$ always include cross-prediction tasks.

The five investigated reservoir topologies are summarized in Fig.~\ref{fig:network_diagram} for a small example network of 32 neurons (to visualize different topologies clearly). They are constructed such that the connections {\em and} their weights can be modified independently. This is achieved by taking the quadratic reservoir matrix $W$ as a Hadamard product, which is element-wise multiplication, of a connection matrix $A$ and a weight matrix $W^c$, i.e., $W= A \odot W^c$. A similar setup was done in Ref.~\onlinecite{Rathor2025}. For nodes with indices $i$ and $j$, the fixed network connections, $i\to j$ and $j\to i$, are encoded in the connection matrix $A$, and their weights, $w_{ij}$ and $w_{ji}$, in the weight matrix $W^c$. The weights $w_{ij} \in \mathbb{R}$ are sampled from a uniform distribution, $w_{ij}\sim {\cal U}([-0.5, 0.5])$. Thus, $W_{ij} = A_{ij} w_{ij}$. The spectral radius of the reservoir matrix $ W $ is then set to a value $\rho$ by normalizing $W$. The connection matrix $A$ is symmetric if $A = A^{T}$; otherwise, it is asymmetric. For the weight matrix $W^c$, symmetry and asymmetry are defined in the same way. 

The most general topology is a random-asymmetric network (R-A) with $A$ and $W^c$ both asymmetric, which implies unidirectional connections, $W_{ij}\ne 0$, but $W_{ji}=0$, along with bidirectional connections with $W_{ij}\ne 0$ and $W_{ji} \ne 0$. For example, nodes 1 and 6 have a unidirectional connection, while nodes 2 and 5 are bidirectionally connected in Fig.~\ref{fig:network_diagram}. The second topology is a symmetrized connection matrix $A$ in combination with an asymmetric weight matrix $W^c$, i.e., RS-A. The third topology, RS-S, is obtained by taking both $A$ and $W^c$ symmetric, i.e., any pair of nodes is bidirectionally connected. Finally, a (random) Watts-Strogatz (WS) network~\cite{Watts1998} with rewiring probability $p = 1$ is used, which has by definition a symmetric $A$, but can obey either an asymmetric or a symmetric weight matrix $W^c$. This defines WS-A and WS-S in Fig.~\ref{fig:network_diagram}. 

\begin{figure}
	\includegraphics[scale=0.85]{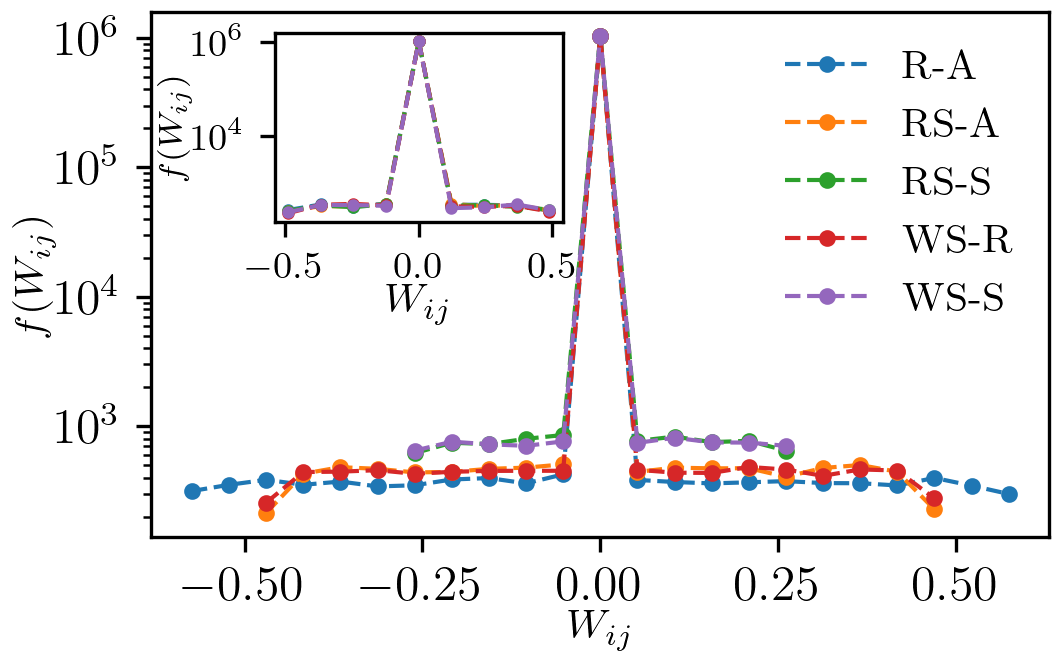}
	\caption{A representative weight distribution $f(W_{ij})$ of reservoir matrix $W$ for the five network configurations, namely, R-A, RS-A, RS-S, WS-A and WS-S with $N=1024$ nodes and reservoir density $D_r = 0.008$ after setting the spectral radius (here, $\rho = 1$). Inset: The weight distributions of the same networks before setting the spectral radius. The pronounced peak at $W_{ij} = 0$ is due to small reservoir density $D_r = 0.008$.}
	\label{fig:weight_dist}
\end{figure}

The realization of these networks starts with the similar distribution of $W_{ij}$, called $f(W_{ij})$, in the reservoir matrix $W$, as shown in the inset of Fig.~\ref{fig:weight_dist}. The distribution $f(W_{ij})$ of the functional reservoir matrix $W$ after setting the spectral radius $\rho = 1$ for all network topologies is shown in Fig.~\ref{fig:weight_dist}. The peak in the distribution corresponds to a large number of null entries in $W$, owing to the small reservoir density of $D_r = 0.008$ which is constant throughout the study. 
 
%--------------------------------------------------------
%--------------------------------------------------------

\section{Dynamical Systems}
The dynamical systems and their configurations of interest for processing via RC are summarized in Table~~\ref{tab:systems_summary}. These are the Mackey-Glass system (MG),~\cite{Mackey_1977} Lorenz 63 system (L63) with $N_{\rm DoF}=3$ degrees of freedom,~\cite{Lorenz1963} the extended Lorenz system (L8) with 8 degrees of freedom,~\cite{Gluhovsky2002} and a Galerkin model for a three-dimensional plane shear flow (SF) with 9 degrees of freedom. \cite{Moehlis_2004} The models L63 and L8 can be derived from a thermal convection flow. The prediction of these systems is performed in open-loop operation of the RC model wherein one-step ahead prediction is realized by feeding the input regularly at each step. Moreover, the fed input is partial with $N_{\rm in} < N_{\rm out}$ dynamical variables, where $N_{\rm out}$ defines the number of target states to predict. 

All of these systems show chaotic dynamics, with a positive maximal Lyapunov exponent $\lambda_{\rm max}$, in their state space. The Kaplan-Yorke dimension $D_{\rm KY}$ provides an estimate of the dimension of the attractor.~\cite{Kaplan_1979} It is given by $D_{\rm KY} = s - \sum_{r=1}^{s}\lambda_r/\lambda_{s+1}$ with $r$th Lyapunov exponent $\lambda_r$ in the Lyapunov spectrum of decreasing Lyapunov exponents such that $\sum_{r=1}^{s}\lambda_r \geq 0$ and $\sum_{r=1}^{s+1}\lambda_r < 0$. The Lyapunov spectrum of a dynamical system is obtained by $\lambda_r = \frac{1}{K} \sum_{k = 1}^K \ln G_r^k $, where $G_r^k$ is the growth rate of the $r$th unit vector in the orthogonal basis of the state space of the dynamical system at time step $k$, and the total number of time steps $K$ is large. The growth rates $G_{r}^{k}$ for $r = 1, 2, ..., N_{\rm DoF}$ are computed using QR-decomposition of the Jacobian matrix $J^k$,~\cite{Geist_1990} which is obtained by simultaneously evolving the state of the system and the dynamics of the corresponding tangent space from an initial condition for the $k$ time steps to iterate the computation of the Jacobian matrix $J^k$ at each time step $k$.

For SF, $D_{\rm KY}$ is computed within the turbulent regime of the shear flow, well before the dynamics becomes laminar eventually. In the following, we will detail all 4 models for completeness. We see that $D_{\rm KY}$ increases from L63 to SF, indicating the increasing complexity of the dynamics in the state space.
\begin{table*}
\centering
\begin{tblr}{cccc}
\hline\hline
Mackey-Glass equation &  Lorenz 63 model & Lorenz-type model & Shear flow model\\ [0.1ex]
(MG) & (L63) & (L8) & (SF)\\ \hline
$N_{\rm DoF}=\infty$ & $N_{\rm DoF}=3$ & $N_{\rm DoF}=8$ & $N_{\rm DoF}=9$\\ [0.1ex]
					  
$N_{\rm in}=1$ & $N_{\rm in}=1$ & $N_{\rm in}=2$ & $N_{\rm in}=5$\\ [0.1ex]
			    & ($A_1$ or $B_1$)&  ($A_1, B_1$) & ($a_1, a_2, a_3, a_4, a_6$)\\ [0.1ex] 
                
$N_{\rm out}=1$ & $N_{\rm out}=3$ & $N_{\rm out}=8$  & $N_{\rm out}=9$\\ [0.1ex]
                & (all) & (all) & (all) \\
$\lambda_{\rm max} = 0.006$ & $\lambda_{\rm max} = 0.91$ & $\lambda_{\rm max} = 1.48$ & $\lambda_{\rm max} = 0.02$ \\
 $D_{\rm KY} = 2.10$ & $D_{\rm KY} = 2.06$ & $D_{\rm KY} = 3.44$ & $D_{\rm KY} = 6.25$ \\
{\includegraphics[scale=0.8]{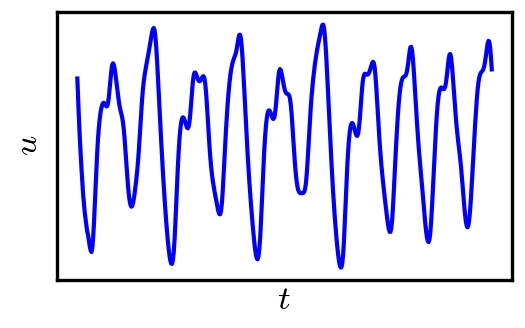}} & \includegraphics[scale=0.8]{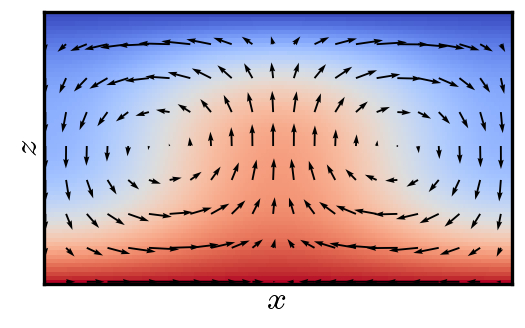} & \includegraphics[scale=0.8]{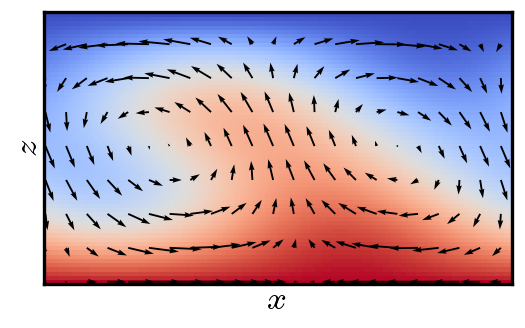} & \includegraphics[scale=0.45]{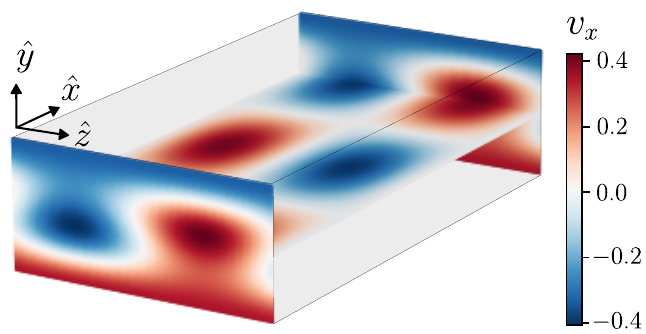}\\
\hline\hline
\end{tblr}        
\caption{Summary of the dynamical systems together with their number of degrees of freedom $N_{\rm dof}$, the numbers of inputs $N_{\rm in}$ and outputs $N_{\rm out}$, as well as their maximal Lyapunov exponent $\lambda_{\rm max}$ and Kaplan-Yorke dimension $D_{\rm KY}$. These are, from left to right, the Mackey-Glass equation (MG) given by Eq.~\eqref{eq:MG17}, the Lorenz 63 and 8-dimensional Lorenz-type models (L8), which are given by Eqs.~\eqref{eq:L8}, and 9-dimensional Galerkin model for a three-dimensional plane shear flow (SF), which is given by  Eqs.~\eqref{eq:L9}. All prediction tasks for the temporal dynamics of these nonlinear models are performed in the open-loop reservoir computing scenario with the listed input and output degrees of freedom (DoF). In the last row, we display a time series (MG) or a snapshot at a time instant (L63, L8, SF) that shows the flow structures. In L63 and L8, these are the temperature field (colored background contours) together with the velocity vector field. In case of SF, these are contours in three planes, displaying the three-dimensional spatial structure of the streamwise velocity field component.}
\label{tab:systems_summary}
\end{table*}

\subsection{Mackey--Glass equation}
The first system we investigate is the nonlinear time-delayed Mackey-Glass (MG) equation~\cite{Mackey_1977} and is given by
\begin{equation}
	\frac{du}{dt} = a \frac{u(t-\tau)}{1+u(t-\tau)^q} - b u(t)\,.
	\label{eq:MG17}
\end{equation}
We use standard parameters $a = 0.2$, $b = 0.1$, $q = 10$, and the time delay $\tau=17$. The dynamics described by time-delayed equation \eqref{eq:MG17} for the parameters, which is chaotic in an attractor with box counting dimension $2<D<3$,~\cite{Ziessler2019} happens in an infinite-dimensional state space. Although the MG equation represents only a scalar function $u(t)$, it is highly nonlinear and exhibits high dynamical variability that depends on the delay parameter $\tau$.

The Mackey--Glass time series~\cite{Mackey_1977} is obtained by iterating the discrete approximation of the delay differential equation \eqref{eq:MG17} by \citet{Grassberger1983}. The initial conditions are sampled from a random uniform distribution, and the first $250,000$ iterations are discarded as initial transients. The time step of the Grassberger-Procaccia iteration is $ \delta t=1.7 \times 10^{-2} $. Further, training and testing data sets are prepared by sampling the time series with a step size of $\Delta t=1$ and rescaling them such that $u(t)\in [-1,1]$. The task is to make a one-step ahead direct prediction of MG time series using RC, see also Table~\ref{tab:systems_summary}.

\subsection{Lorenz 63 model and extended 8-dimensional Lorenz model}\label{rbc}
\begin{figure}
	\includegraphics[scale=0.45]{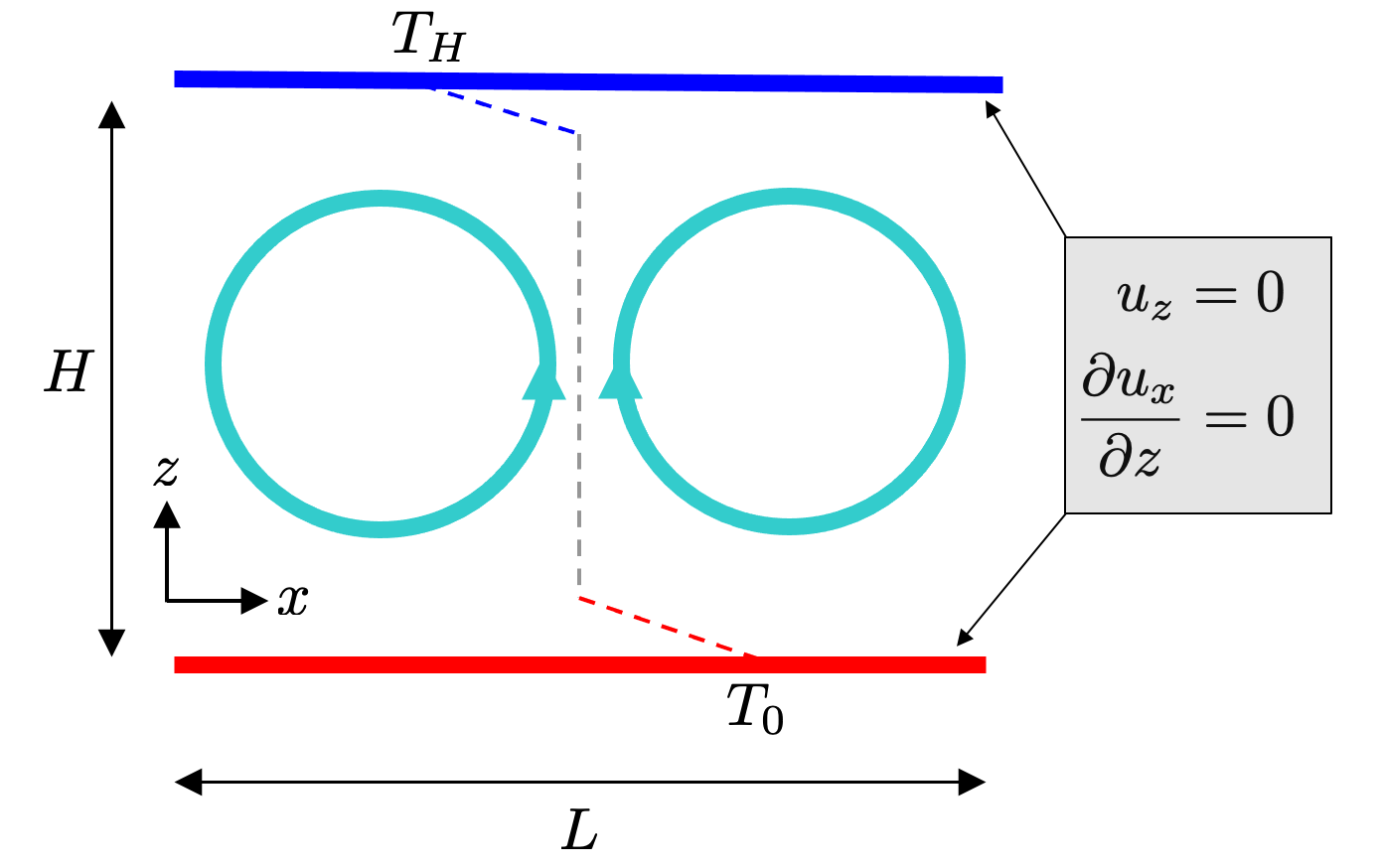}
	\caption{Sketch of the two-dimensional Rayleigh-B\'{e}nard convection configuration for both Lorenz models. We indicate the boundary conditions and the counter-rotating flow circulations in the form of convection rolls in the middle of the layer.}
	\label{fig1}
\end{figure}

The standard three-dimensional Lorenz model, which is often referred to as the Lorenz 63 model~\cite{Lorenz1963} as well as its extension to 8 dimensions,~\cite{Tong2002,Gluhovsky2002} can be derived from the equations for buoyancy-driven two-dimensional Rayleigh-B\'{e}nard convection flow between two impermeable parallel plates at distance $H$.~\cite{Chilla2012} The configuration together with the corresponding boundary conditions is shown in Fig.~\ref{fig1}. A fluid flow layer between two impermeable plates is heated uniformly from below and cooled from above. The top plate is kept at a lower temperature $T_H$ than the bottom plate, which is at $T=T_0$. Thus, $\Delta T=T_0-T_H>0$. At the top and bottom, free-slip boundary conditions (see the gray box to the right in the figure) are set for the two-dimensional incompressible velocity vector field ${\bm u}=(u_x(x,z,t),u_z(x,z,t)$. In $x$-direction periodic boundary conditions are applied. The equilibrium state of the convection layer is given by ${\bm u}=0$ and $T_{\rm eq}(z)=T_0-z \Delta T/H$. Once the temperature difference $\Delta T$ exceeds a critical threshold value, fluid motion in the form of the sketched circulation rolls starts by an onset of a linear flow instability. The (turbulent) fluid motion is driven by buoyancy forces.

%--------------------------------------------------------
%----------------------------------------------------------

All Lorenz-type models are obtained by spatial mode expansions of the scalar velocity stream function $\zeta(x,z,t)$ (which substitutes the incompressible two-dimensional velocity vector field) and the temperature fluctuation field $\theta(x,z,t)=T(x,z,t)-T_{\rm eq}(z)$. These expansions satisfy the boundary conditions, which we outlined in the caption of Fig.~\ref{fig1}. The series expansions are given by

\begin{align}
	\zeta(x,z,t)=\sum_{i,j=1}^\infty c_\zeta A_{ij}(t) \Phi(\alpha_i x) \sin(\beta_j z)\,,	\label{exp1}\\
	\theta(x,z,t)=\sum_{k,l=1}^\infty c_\theta B_{kl}(t) \Phi(\alpha_k x) \sin(\beta_l z)\,, \label{exp2}
\end{align}

with normalization prefactors $c_\psi , c_\theta$, the real amplitudes $\{A_{ij}(t),B_{kl}(t)\}$, and the wavenumbers $\alpha_k=k \alpha=2\pi k/\Gamma$ and $\beta_k=k\beta=k\pi$. In the extended Lorenz model, truncation is done after the fourth term in both expansions, resulting in 8 modes. We take {the critical wavelength for linear instability} $\Gamma=L/H=2\sqrt{2}$ with periodicity length $L$ { for $b=8/3$ in Eq.~(\ref{eq:B2})}.  

The four selected velocity modes are $A_1=A_{11},\, A_2=A_{01},\, A_3=A_{12}$ and $A_4=A_{03}$; the four selected temperature modes are $B_1=B_{11},\, B_2=B_{02},\, B_3=B_{12}$ and $B_4=B_{04}$ in accordance with the notation in Eqs.~\eqref{exp1} and \eqref{exp2}. The set of coupled nonlinear ordinary differential equations, our 8-dimensional Lorenz system (L8), is given by
\begin{widetext}
\begin{subequations}\label{eq:L8}
\begin{align}
	\frac{dA_1}{d\tau}&=\sigma \left(B_1 - A_1\right) - \frac{c_3}{\sqrt{2}c_1} A_2A_3 + \frac{3c_{-5}}{\sqrt{2}c_1} A_3A_4\,, \label{eq:A1} \\
	\frac{dB_1}{d\tau}&=-B_1 + rA_1 - A_1B_2 + \frac{1}{2}A_2B_3 +  \frac{3}{2}A_4B_3\,,\label{eq:B1}\\
	\frac{dB_2}{d\tau}&= - bB_2 + A_1 B_1 \,,\label{eq:B2}\\
	\frac{dA_2}{d\tau}&=-\frac{\sigma b}{4} A_2 -\frac{3}{2\sqrt{2}} A_1A_3\,,\label{eq:A2}\\
	\frac{dA_3}{d\tau}&=-\sigma \left(\frac{c_4}{c_1} A_3 + \frac{c_1}{\sqrt{2}(c_4)}B_3\right) + \frac{c_0}{\sqrt{2}c_1} A_1A_2 - \frac{3c_{-8}}{\sqrt{2}(c_4)} A_1A_4\,,\label{eq:A3}\\
	\frac{dA_4}{d\tau}&=-\frac{9 \sigma b}{4} A_4-\frac{1}{2\sqrt{2}} A_1A_3\,,\label{eq:A4}\\
	\frac{dB_3}{d\tau}&= -\frac{c_4}{c_1}B_3 -A_2B_3 + \sqrt{2}r A_3 + 3A_4B_1 - 2\sqrt{2} A_3B_4\,,\label{eq:B3}\\
	\frac{dB_4}{d\tau}&=-4b B_4 + \frac{3\sqrt{2}}{4} A_3B_3\,,\label{eq:B4}
\end{align}
\end{subequations}
\end{widetext}
where $c_n = \alpha^2 + n\beta^2$ ($n \in \mathbb{Z}$), Prandtl number $\sigma = 10$, $b=8/3$, and relative Rayleigh number $r=28$. The rescaled time is $\tau=c_1t$. This model is integrated numerically with a 4th-order Runge-Kutta scheme and the obtained time series of the expansion coefficients are used as input as well as ground truth in the reservoir computing experiments. The integration time step is $\delta t = 2 \times 10^{-3}$. The initial conditions are randomly chosen, and a long enough transient at the beginning is discarded. The additional higher-order modes with their much smaller amplitudes lead to a conservation of total energy $E$ and vorticity $\Omega$. Both invariants are given by 
\begin{equation}
	E(t)=\frac{1}{2A}\int_A ((\nabla\zeta)^2-z\theta) dA\quad \mbox{and}\quad
	\Omega(t)=\frac{1}{A}\int_A \omega dA\,,   
\end{equation}
with the vorticity $\omega=-\nabla^2\zeta$ and the convection domain size $A=LH$. 

The well-known Lorenz 63 model includes $A_1$, $B_1$, and $B_2$ only.~\cite{Lorenz1963} The data for L63 model is also generated similar to L8 model with the same integration time step of $\delta t = 2 \times 10^{-3}$. The flow constructed from the modes of extended Lorenz model describes the shear motions as shown by the tilt in velocity and temperature fields. A snapshot of the convection flow dynamics for L63 and L8 is shown in Table~\ref{tab:systems_summary}.

\subsection{Galerkin model for plane shear flow}\label{sf}
The third dynamical system, which we use in our investigation, is a 9-mode Galerkin model~\cite{Waleffe1997,Moehlis_2004} of a simple shear flow, sometimes also known as the minimal flow unit.\cite{Hamilton1995} This three-dimensional nonlinear model describes the elementary dynamical cycle of near-wall coherent flow structures present in every wall-bounded turbulent flow. The present model is for a shear flow between parallel walls at distance $d_0$ with free-slip boundary conditions for the velocity field subject to a sinusoidal body force which is added to the Navier-Stokes equations to sustain the sinusoidal shear flow \cite{Moehlis_2004} in the domain of $0 \le x \le L_x$, $-1 \le y \le 1$, and $0\le z \le L_z$.  The starting point is the expansion of velocity field ${\bm v}(x,y,z,t) = (v_x, v_y, v_z)$ as 
\begin{equation}
	{\bm v}(x,y,z,t) = \sum_{i = 1}^{9} a_i(t) {\bm v}_i(x,y,z)\,,
	\label{eq:SF}
\end{equation}
with prescribed spatial modes ${\bm v}_i$, which all satisfy the free-slip boundary conditions with respect to the $y$ or wall-normal direction and periodicity in the other two space directions. The detailed structure of the modes is found in Ref.~\onlinecite{Moehlis_2004}. The modes considered are the basic sinusoidal shear flow profile ${\bm v}_1(y)=\sqrt{2}\sin(\pi y/2) {\bm e}_x$, the streamwise streak ${\bm v}_2$, the downstream vortex ${\bm v}_3$, the spanwise flow modes ${\bm v}_4$ and ${\bm v}_5$, the normal vortex modes ${\bm v}_6$, ${\bm v}_7$, and ${\bm v}_8$, and finally the modification of the basic velocity profile due to turbulence ${\bm v}_9$. We take $L_x = 4\pi$ and $L_z = 2\pi$.

The dynamics of these modes is obtained by a projection of the three-dimensional Navier-Stokes dynamics on each of the 9 modes. This results in a coupled set of nonlinear ordinary differential equations for the expansion coefficients $a_i(t)$ in Eq.~\eqref{eq:SF} when higher-order nonlinearities are truncated. The time evolution of the coefficients $a_i(t)$ with $i=1,...,9$, our dynamical system SF, is given by
\begin{widetext}
\begin{subequations}\label{eq:L9}
\begin{align}
\frac{da_1}{dt} &= \frac{\beta^2}{\rm Re}(1-a_1) - \sqrt{\frac{3}{2}} \frac{\beta \gamma}{K_{\alpha\beta\gamma}} a_6 a_8 + \sqrt{\frac{3}{2}} \frac{\beta \gamma}{K_{\beta \gamma}} a_2 a_3\,, \label{eq:L9_a1}\\
\frac{da_2}{dt} &= -\left(\frac{4\beta^2}{3} + \gamma^2\right)\frac{a_2}{\textrm{Re}} + \frac{\gamma^2}{\sqrt{6}\kappa_{\alpha\gamma}} \left( \frac{10}{3}a_4 a_6 - a_5 a_7 \right)-\frac{\alpha\beta\gamma}{\sqrt{6}\kappa_{\alpha\gamma}\kappa_{\alpha\beta\gamma}} a_5 a_8 - \sqrt{\frac{3}{2}} \frac{\beta \gamma}{\kappa_{\beta\gamma}} \left( a_1 a_3 + a_3 a_9\right)\,, \label{eq:L9_a2}\\
\frac{da_3}{dt} &= -\frac{\beta^2 + \gamma^2}{\textrm{Re}}a_3 + \frac{2}{\sqrt{6}} \frac{\alpha\beta\gamma}{\kappa_{\alpha\gamma}\kappa_{\beta\gamma}}(a_4 a_7 + a_5 a_6) + \frac{\beta^2 (3\alpha^2 + \gamma^2) - 3\gamma^2 (\alpha^2 + \gamma^2)}{\sqrt{6}\kappa_{\alpha\gamma}\kappa_{\beta\gamma}\kappa_{\alpha\beta\gamma}} a_4 a_8\,, \label{eq:L9_a3}\\
\frac{da_4}{dt} &= -\frac{3\alpha^2 + 4\beta^2}{3\textrm{Re}}a_4 - \frac{\alpha}{\sqrt{6}} a_1 a_5 - \frac{10}{3\sqrt{6}} \frac{\alpha^2}{\kappa_{\alpha\gamma}} a_2 a_6 -\sqrt{\frac{3}{2}} \frac{\alpha\beta\gamma}{\kappa_{\alpha\gamma}\kappa_{\beta\gamma}} a_3 a_7 - \sqrt{\frac{3}{2}} \frac{\alpha^2 \beta^2}{\kappa_{\alpha\gamma}\kappa_{\beta\gamma}\kappa_{\alpha\beta\gamma}} a_3 a_8 - \frac{\alpha}{\sqrt{6}} a_5 a_9\,, \label{eq:L9_a4}\\
\frac{da_5}{dt} &= - \frac{\alpha^2 + \beta^2}{\Re} a_5 + \frac{\alpha}{\sqrt{6}} a_1 a_4 + \frac{\alpha^2}{\sqrt{6}\kappa_{\alpha\gamma}} a_2 a_7 - \frac{\alpha\beta\gamma}{\sqrt{6} \kappa_{\alpha\gamma}\kappa_{\alpha\beta\gamma}} a_2 a_8 + \frac{\alpha}{\sqrt{6}} a_4 a_9 + \frac{2}{\sqrt{6}} \frac{\alpha \beta \gamma}{\kappa_{\alpha\gamma}\kappa_{\beta\gamma}} a_3 a_6\,, \label{eq:L9_a5}\\
\frac{da_6}{dt} &= -\frac{3\kappa_{\alpha\beta\gamma}^2 + \beta^2}{3\Re} a_6 + \frac{\alpha}{\sqrt{6}} a_1 a_7 \sqrt{\frac{3}{2}} \frac{\beta\gamma}{\kappa_{\alpha\beta\gamma}}a_1 a_8 + \frac{10}{3\sqrt{6}} \frac{\alpha^2 - \gamma^2}{\kappa_{\alpha\gamma}} a_2 a_4 -2 \sqrt{\frac{2}{3}} \frac{\alpha\beta\gamma}{\kappa_{\alpha\gamma}\kappa_{\beta\gamma}} a_3 a_5\nonumber \\  
& \phantom{=} +\frac{\alpha}{\sqrt{6}} a_7 a_9 + \sqrt{\frac{3}{2}} \frac{\beta \gamma}{\kappa_{\alpha\beta\gamma}} a_8 a_9\,, \label{eq:L9_a6}\\
\frac{da_7}{dt} &= -\frac{\kappa_{\alpha\beta\gamma}^2}{\Re} a_7 - \frac{\alpha}{\sqrt{6}} (a_1 a_6 + a_6 a_9) + \frac{1}{\sqrt{6}} \frac{\gamma^2 - \alpha^2 }{\kappa_{\alpha\gamma}} a_2 a_5 + \frac{1}{\sqrt{6}} \frac{\alpha\beta\gamma}{\kappa_{\alpha\gamma}\kappa_{\beta\gamma}}a_3 a_4\,, \label{eq:L9_a7}\\
\frac{da_8}{dt} &= -\frac{\kappa_{\alpha\beta\gamma}^2}{\Re} a_8 + \frac{2}{\sqrt{6}} \frac{\alpha\beta\gamma}{\kappa_{\alpha\gamma}\kappa_{\alpha\beta\gamma}} a_2 a_5 + \frac{\gamma^2 (3\kappa_{\alpha\gamma}^2 - \beta^2)}{\sqrt{6}\kappa_{\alpha\gamma}\kappa_{\beta\gamma}\kappa_{\alpha\beta\gamma}} a_3 a_4\,, \label{eq:L9_a8}\\
\frac{da_9}{dt} &= -\frac{9\beta^2}{\Re}a_9 + \sqrt{\frac{3}{2}} \frac{\beta\gamma}{\kappa_{\beta\gamma}} a_2 a_3 - \sqrt{\frac{3}{2}} \frac{\beta\gamma}{\kappa_{\alpha\beta\gamma}} a_6 a_8\,, \label{eq:L9_a9}
\end{align}
\end{subequations}
\end{widetext}
where $\kappa_{\alpha\gamma} = \sqrt{\alpha^2 + \gamma^2}$, $ \kappa_{\beta\gamma}  = \sqrt{\beta^2 + \gamma^2}$, and $ \kappa_{\alpha\beta\gamma} = \sqrt{\alpha^2 + \beta^2 + \gamma^2} $. Also $\alpha=2\pi/L_x$, $\beta=\pi/2$, and $\gamma=2\pi/L_z$. The central parameter in the dynamical model is the Reynolds number $\Re=U_0 d_0/(2\nu)$. Here, $U_0$ is the characteristic amplitude of the basic flow and $\nu$ is the kinematic viscosity of working fluid. Table \ref{tab:systems_summary} shows a snapshot of the spatial structure of the streamwise velocity component $v_x$ reconstructed from ansatz \eqref{eq:SF} displaying the typical streamwise streaky structures.

The data is generated by numerically integrating the SF model in Eqs.~\ref{eq:L9} for Re = 500 using a Runge-Kutta-Fehlberg (RK45) scheme with initial conditions selected randomly on an energy shell $\sum_{i=1}^{9} a_i^2 = 0.1$. The integration time step is $2\times 10^{-3}$. This completes the description of the 4 dynamical systems which we study in the following. 

%---------------------------------------------------------------------
\begin{figure*}
	\includegraphics[scale=0.8]{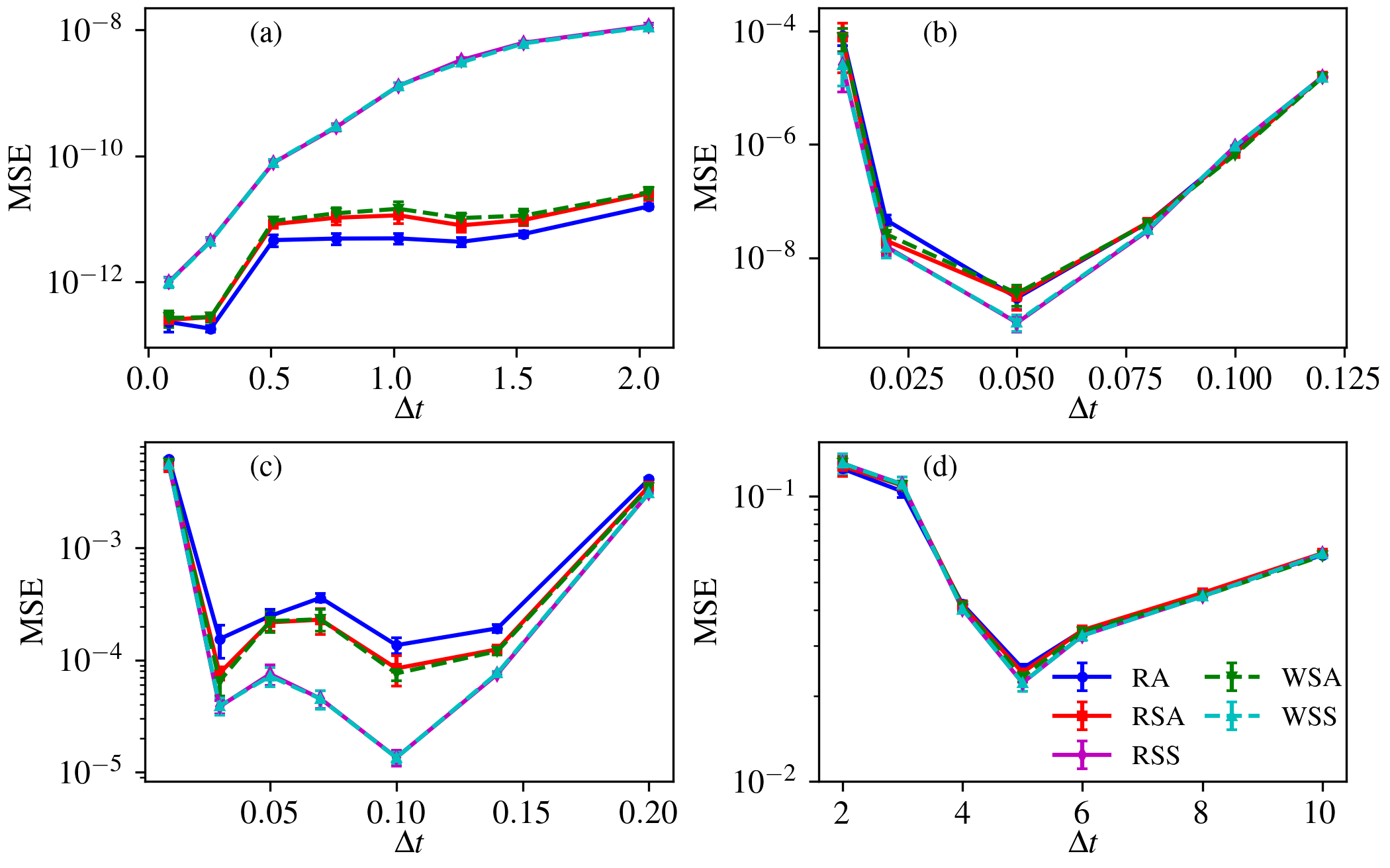}
	\caption{Optimal time step size for sampling the input data of the (a) Mackey--Glass equation (MG), (b) Lorenz 63 (L63) model, (c) 8-dimensional Lorenz-type (L8) model, and (d) Galerkin model of the three-dimensional plane shear flow (SF) for reservoir networks with node numbers $N=1024$. Each case for each application, and each topology is obtained as an ensemble median of 50 different random reservoir network initializations. The legend in (c) applies to all panels.}
	\label{fig:optimum_sample_step_size}
\end{figure*}
%---------------------------------------------------------------------

\section{Results}
In the following, we compare the performance of the five different reservoir computing models, each applied to the prediction tasks in 4 different nonlinear dynamical systems, MG, L63, L8, and SF, which we described in detail in Sec. III. The input data from all the systems is rescaled to an interval of [-1, 1]. The five RC models differ in the topology of their random reservoirs, which we have detailed in Sec. II, namely the reservoir topologies R-A, RS-A, RS-S, WS-A, and WS-S.    

\subsection{Optimal reservoir computing time step}\label{subsec:rc_time_step}
The reservoir update proceeds in correspondence with Eq.~\eqref{eq:RC} with a time step $\Delta t$ to advance from time instant $k$ to the next, $k+1$ in this notation. This time step $\Delta t$ defines the sampling time step and is a multiple of the actual numerical integration time step of the dynamical system, the latter of which we denoted by $\delta t$. RC models respond differently to the sampling of input data as shown in Ref.~\onlinecite{Jaurigue2024}. Thus, each RC model in combination with the underlying task has a characteristic time step to perform the computation optimally. To this end, the reservoir computing time step, i.e., the sampling interval of input data $\Delta t$, is determined first, by optimizing the median MSE, in Eq.~(\ref{eq:MSE}), of each DPT over an ensemble of 50 randomly initialized reservoirs for each topology. The number of neurons has been fixed to $N=1024$ for this pre-analysis. The DPTs correspond to those, which are indicated in Table~\ref{tab:systems_summary}. 

\begin{figure*}
	\includegraphics[scale=0.8]{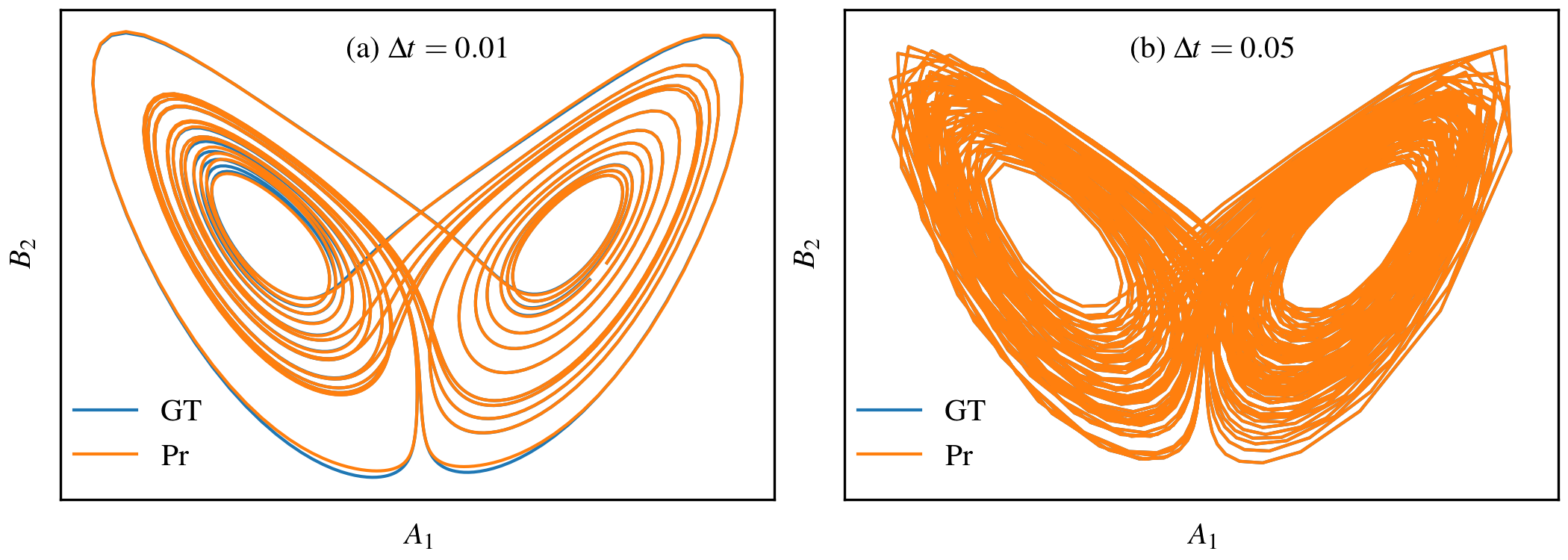}
	\caption{Reconstruction of the attractor of Lorenz 63 system with respect to output modes $A_1$ and $B_2$ for reservoir computing time step (a) $\Delta t = 0.01$ and (b) $\Delta t = 0.05$. The input mode is $B_1$. Solid blue line shows ground truth (GT) of $A_1$ and $B_1$, and solid orange line represents their prediction (Pr).}
	\label{fig:attractor_L63}
\end{figure*}

Figure~\ref{fig:optimum_sample_step_size} summarizes the results on the reservoir computing time step for open-loop scenario with $N_{in} < N_{out}$ (see Table ~\ref{tab:systems_summary}). The integration time step for DPTs, namely L63, L8 and SF, is $\delta t  = 0.002$. The data is sampled at an interval of $\Delta t = 25 \delta t = 0.05$ for L63, $\Delta t = 50 \delta t = 0.1$ for L8, and $\Delta t = 2500 \delta t = 1$ for SF. It is observed that there is practically no influence of reservoir topology on the MSE in the cases of L63 and SF, see Fig.~\ref{fig:optimum_sample_step_size}(b) and (d). For the 8-dimensional extended Lorenz model, a dependence of the MSE on the topology is observed in Fig.~\ref{fig:optimum_sample_step_size}(c). The optimal reservoir computing time step remains practically the same for all reservoir topologies. The manifestation of the optimal sampling time step is illustrated in Fig.~\ref{fig:attractor_L63}, which shows the attractor of L63 with modes $A_1$ and $B_1$ reconstructed by RC with input mode $B_1$ for $\Delta t = 0.01$ and the optimal $\Delta t = 0.05$.  The reconstruction of the attractor with input data $B_1$ sampled at $\Delta t = 0.05$ is visually better than that at $\Delta t = 0.01$, although the former is coarser and no longer smoother compared to the latter. This indicates that RC learns better with coarser input data and enables longer prediction. 

Finally, in the case of the MG system, we find that the smaller the reservoir computing time step $\Delta t$, the better the prediction, see Fig.~\ref{fig:optimum_sample_step_size}(a). We work for the following analysis with $\Delta t=1$. After preparing the input data, we can now proceed to a detailed performance analysis in the next Subsection.

%---------------------------------------------------------------------
%---------------------------------------------------------------------

\subsection{Performance for dynamics prediction task}
In the following, we present the performance of all reservoir network topologies for their optimal hyperparameter sets. All DPTs carried out using reservoirs of $N = 1024$ nodes with a small reservoir density of $D_r = 0.008$ are thoroughly explored and discussed in detail. In addition, we consider smaller reservoir sizes with $N= 256$ and 512 nodes for comparison. The optimal hyperparameters---spectral radius $\rho$, leaking rate $\varepsilon$, and regularization coefficient $\gamma$---are determined on a three-dimensional parameter grid to minimize the median of the mean square error (MSE) in Eq.~\eqref{eq:MSE} over different realizations of the reservoir coupling matrix $W$ for a given and fixed input weight matrix $W^{\rm in}$ for each $N$. The mean square error between the ground truth vector $\mb{y}(k) = [y_1, y_2, ..., y_{N_{out}}]^T$ and the predicted vector $\mb{y}^p(k) = [y^p_1, y^p_2, ..., y^p_{N_{out}}]^T$ is obtained by
%-------------------------------
 \begin{equation}
 	{\rm MSE} = \frac{1}{k_p - k_t}\sum_{k=k_t + 1}^{k_p} \left \lVert\mb{y}(k) - \mb{y}^p(k)\right\rVert^2\,,
 	\label{eq:MSE}
 \end{equation}
%----------------------------
where $\| \cdot\|$ denotes $L^2$ norm. This can be written as the sum of the MSEs of all $N_{\rm out}$ tasks of a DPT, i.e.,
%-------------------------------
\begin{equation}
	{\rm MSE} = \sum_{d=1}^{N_{\rm out}} {\rm MSE}_{d},
	\label{eq:MSE_sum}
\end{equation}
%----------------------------
where ${\rm MSE}_{d}$ is the prediction error of $d$th task (or component), which is defined as
%-------------------------------
\begin{equation}
	{\rm MSE}_{d} = \frac{1}{k_p - k_t}\sum_{k=k_t + 1}^{k_p} ({y_{d}}(k) - {y^p_{d}}(k))^2\,.
	\label{eq:MSE_task}
\end{equation}
%----------------------------
The analysis of each DPT is performed over an ensemble of $10^3$ different realizations of the reservoir coupling matrix with fixed input weights for all simulations with a given number of nodes $N$.\cite{viehweg_parameterizing_2023} We also analyze the relative improvement of the performance of the MSE for a specific DPT with respect to the network topology that performs the {\em worst} with mean square error MSE$^w$ for the same DPT. This measure is given by
\begin{equation}
 	I_{\rm err} = \frac{\vert  {\rm MSE}^w - {\rm MSE} \vert}{{\rm MSE}^w}.
 	\label{eq:I_err}
\end{equation}
The grid-search is performed in hyperparameter space defined by varying spectral radius in $[0.0, 1.5]$, leaking rate in $[0.1, 1.0]$, and regularization parameter in $[10^{-13}, 1]$ for each reservoir size $N$. The optimal hyperparameters for $N = 1024$ are listed in Table~~\ref{tab:hp_optimized} for each DPT.

%-----------------------------------------------
\begin{table}
	\begin{center}
		\begin{tabular}{lcccc}
			\hline\hline
			& MG & L63 & L8 & SF\\
			& $(\rho,\gamma)$ & $(\rho,\gamma)$ & $(\rho,\gamma)$ & $(\rho,\gamma)$\\
			\hline
			R-A	& $( 1.2, 10^{-11})$ & $( 0.7,10^{-11})$ & $( 0.5,10^{-11})$ & $( 0.9,10^{-1})$\\
			RS-A & $( 1.3,10^{-11})$ & $( 0.8,10^{-11})$ & $( 0.9,10^{-10})$ & $( 1.1,10^{-2})$\\
			RS-S & $( 1.2,10^{-11})$ & $( 1.0,10^{-11})$ & $( 1.0,10^{-11})$ & $( 1.0,10^{-2})$\\
			WS-A & $( 1.3,10^{-11})$ & $( 0.8,10^{-11})$ & $( 0.9,10^{-10})$ & $( 1.1,10^{-2})$\\
			WS-S & $( 1.2,10^{-11})$ & $( 1.0,10^{-11})$ & $( 1.0,10^{-11})$ & $( 1.0,10^{-2})$\\
			\hline\hline
		\end{tabular}
		\caption{Comparison of the optimal reservoir computing model hyperparameters spectral radius $\rho$ and regularization parameter $\gamma$ with fixed leaking rate $\varepsilon = 0.7$ obtained from a grid search in an open-loop prediction scenario. Three learning tasks, Mackey-Glass equation (MG), 3- and 8-dimensional Lorenz model (L63, L8), and 9-dimensional Galerkin model for a three-dimensional shear flow (SF) are analyzed. Five different random network topologies of the reservoir are used for each case. They have been obtained from an ensemble of $10^2$ different initial random realizations of the reservoir matrix $W$ with $N=1024$ nodes.}
		\label{tab:hp_optimized}
	\end{center}
\end{table}
%---------------------------------------------------------------------
\begin{figure*}
	\includegraphics[scale=0.8]{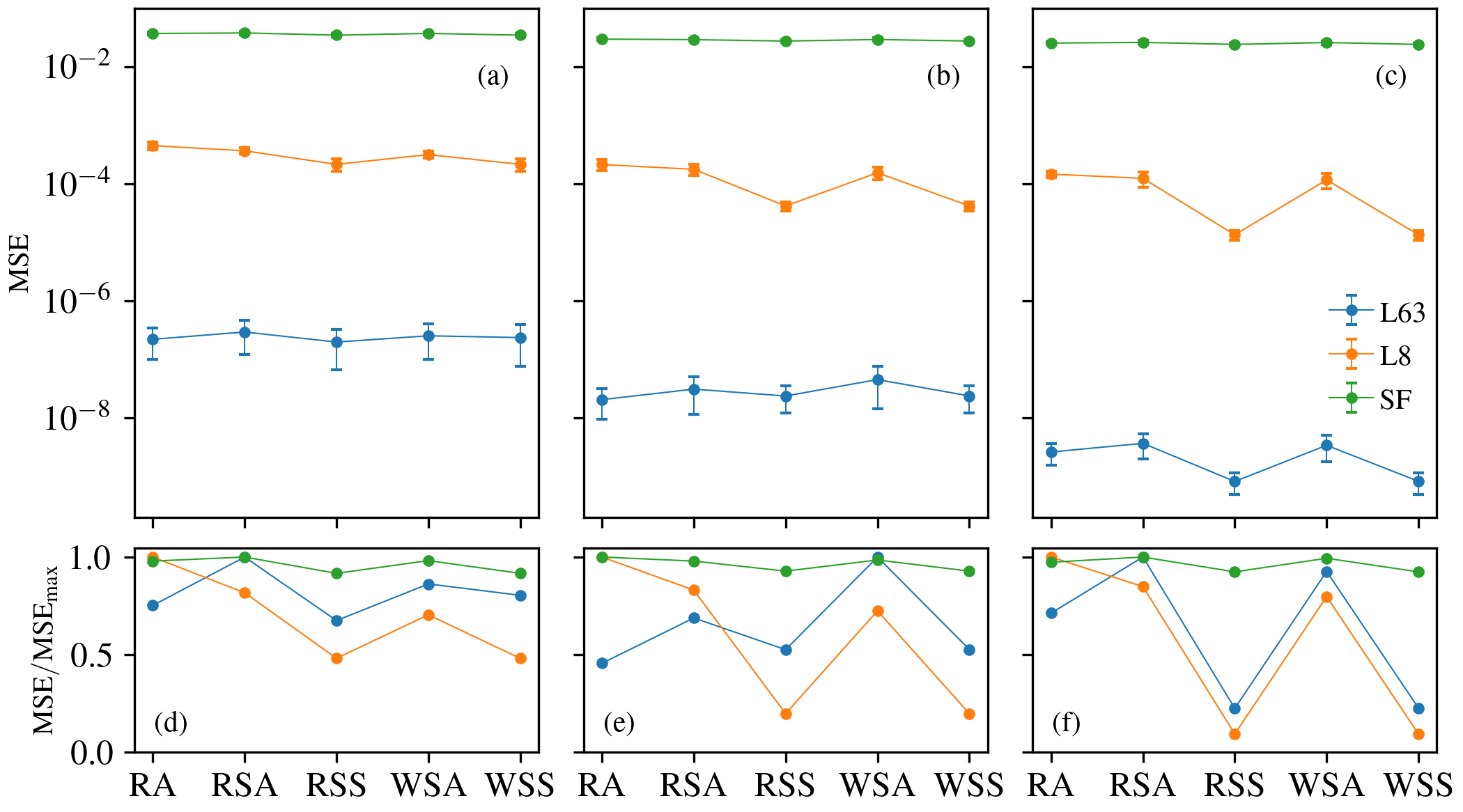}
	\caption{Comparison of the performance of the 5 different reservoir computing models from prediction tasks of the Lorenz 63 (L63), the 8-dimensional Lorenz-type (L8), and the Galerkin model of the three-dimensional plane shear flow (SF) for reservoir networks with node numbers (a) 256, (b) 512, (c) 1024. See also Table~\ref{tab:systems_summary} for details on the number of inputs and outputs. In panels (a--c), we display the mean square error (MSE). Panels (d--f) display the MSE normalized by maximum MSE of each task. Each case for each application, each topology, and each number of reservoirs is obtained as an ensemble median of $10^3$ different random reservoir network initializations. The legend in panel (c) applies to all six panels.}
	\label{fig:mse_comparison}
\end{figure*}
%---------------------------------------------------------------------

The MSE results of the DPTs are summarized in Table~~\ref{tab:mse_summary}. The smallest MSE values are always indicated in bold. For the MG case with $N_{\rm out} = 1$, the random network with R-A topology predicts the best among all the networks studied, which confirms the results of Ref.~\onlinecite{Rathor2025}. The prediction from the networks RS-A and WS-A, which are similar in connection and weight distribution, are comparable. Similar results follow for networks RS-S and WS-S. For the cases of L63 with $N_{\rm out} = 3$ and L8 with $N_{\rm out} = 8$, it is the symmetric reservoir topology for, both connection and weights (namely RS-S and WS-S), that performs best among the five reservoir topologies. For the SF case with $N_{\rm out} = 9$, the performance difference between RS-A (WS-A) and RS-S (WS-S) is significantly reduced. Thus, we observe a trend that symmetric networks perform better dynamics prediction with respect to other network topologies when $N_{\rm in} < N_{\rm out}$. The differences seem, however, to decrease with growing dimensionality.

The robustness of this result is further tested for smaller reservoir sizes of node numbers $N=256$ in panel (a) and $N = 512$ in panel (b) of Fig.~\ref{fig:mse_comparison}. The MSE improves with the reservoir size for all tasks, even though this trend remains small for SF. The trend is also visible when showing the MSE normalized by maximum MSE, $\rm MSE_{\rm max}$, for each task. This is shown in Figs.~\ref{fig:mse_comparison}(d), (e), and (f). In the following, we further investigate the individual tasks and their cumulative effect on the corresponding DPT.

%-----------------------------------------------
\begin{table*}
	\begin{tabularx}{1\textwidth}{ 
			 >{\hsize=.38\hsize\raggedright\arraybackslash}X 
			 >{\hsize=1.15\hsize\centering\arraybackslash}X 
			 >{\hsize=1.15\hsize\centering\arraybackslash}X
			 >{\hsize=1.15\hsize\centering\arraybackslash}X
			 >{\hsize=1.15\hsize\centering\arraybackslash}X
		 }
		\hline\hline
		 & MSE-MG [$I_{\rm err} \%$] & MSE-L63[$I_{\rm err}\%$] & MSE-L8 [$I_{\rm err}\%$] & MSE-SF [$I_{\rm err}\%$]\\
		\hline
    	R-A   & $\bf{(5\pm 1)\times 10^{-12}}$ [99.6] & $(3 \pm 1)\times 10^{-9}$ [25.0]  & $(1.5 \pm 0.2 )\times 10^{-4}$ [0.0] & ${(2.5 \pm 0.1) \times 10^{-2}}$ [3.8] \\
		RS-A  & $(1.6\pm 0.5)\times 10^{-11}$ [98.8] & $(4 \pm 2) \times 10^{-9}$ [0.0] & $(1.2 \pm 0.4) \times 10^{-4}$ [20.0] & $(2.6 \pm 0.2) \times 10^{-2}$ [0.0]  \\
		RS-S & $(1.3\pm 0.2)\times 10^{-9}$ [0.0] & $\bf{(8 \pm 3) \times 10^{-10}}$ [80.0] & $\bf{(1.3 \pm 0.2) \times 10^{-5}}$ [91.3] & $\bf{(2.4 \pm 0.1) \times 10^{-2}}$ [7.7]\\
		WS-A  & $(1.6\pm 0.5)\times 10^{-11}$ [98.8] & $(3 \pm 2) \times 10^{-9}$ [25.0] & $(1.2 \pm 0.4) \times 10^{-4}$ [20.0] & $(2.6 \pm 0.2) \times 10^{-2}$ [0.0]\\
		WS-S  & $(1.3\pm 0.2)\times 10^{-9}$ [0.0] & $\bf{(8 \pm 3) \times 10^{-10}}$ [80.0] & $\bf{(1.3 \pm 0.2) \times 10^{-5}}$ [91.3] & $\bf{(2.4 \pm 0.1) \times 10^{-2}}$ [7.7] \\
		\hline\hline
	\end{tabularx}
	\caption{Performance in an open-loop scenario for Mackey-Glass equation (MG), 3- and 8-dimensional Lorenz-type model (L63, L8), and 9-dimensional Galerkin model of a three-dimensional shear flow (SF). For all cases, we provide the MSE with error bars. The performance improvement $I_{\rm err}$ as defined in Eq.~\eqref{eq:I_err} is specified in parentheses $[\cdot]$, given in per cent. For the three-dimensional shear flow (SF), we list in addition the normalized relative error (NARE) (\%) which is defined in appendix A in Eq.~\eqref{eq:NARE_q}.}
	\label{tab:mse_summary}
\end{table*}

%-----------------------------------------------

\begin{figure*}
	\includegraphics[scale=0.75]{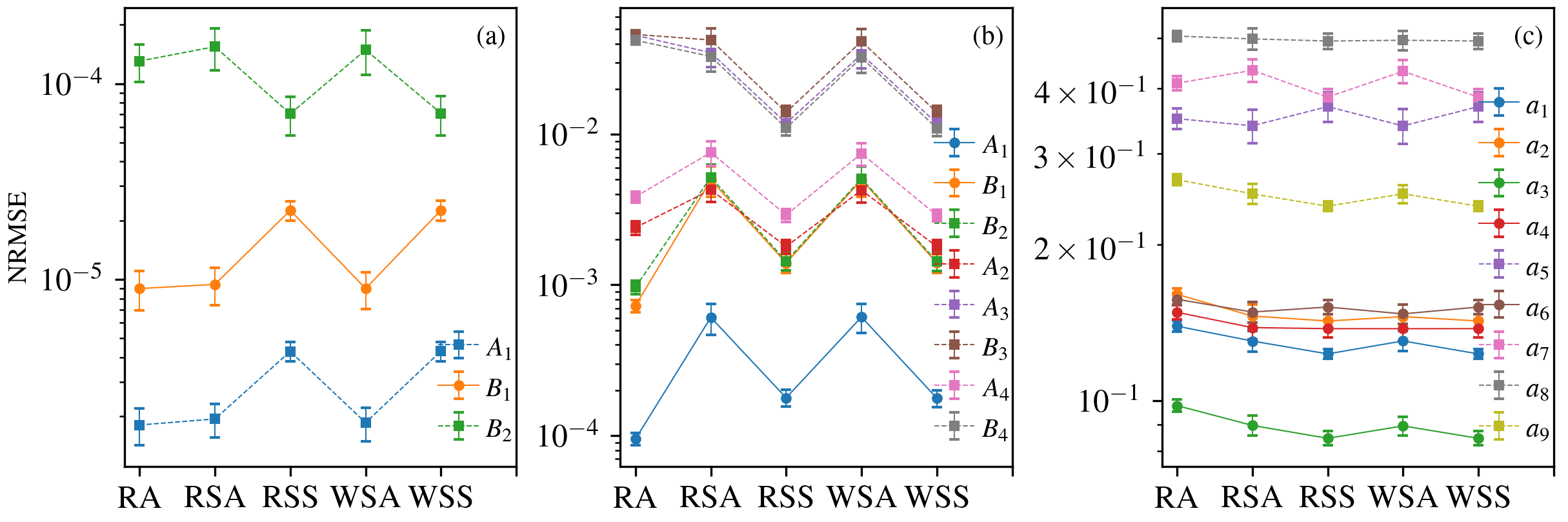}
	\caption{Task-wise comparison of NRMSE, see Eq.~\eqref{eq:NRMSE}, for the network performance of the DPT of (a) L63, (b) L8, and (c) SF for $N = 1024$ nodes. Solid lines with filled circle marker represent direct-prediction tasks; dashed lines with filled square marker show cross-prediction tasks. See Subsections ~\ref{rbc} and ~\ref{sf} for the legends of the three cases. In L63, the mode $B_1$ is fed, in L8 the modes $A_1$ and $B_1$, and in SF the 5 modes given in Table~\ref{tab:systems_summary}.}
	\label{fig:mse_modewise}
\end{figure*}
%----------------------------------------------------

\subsection{Performance for cross-prediction task}
\label{sec:cross-prediction}
In the following, we refine the analysis to component-wise cross predictions to explore the dependence of the performance on symmetry or asymmetry.  In a cross-prediction task, the component $y_d$ is predicted by training the RC model with inputs $u_{d'}$ and $d\ne d'$. In the case of L63, for example, $A_1$ is fed in and $B_1$ is predicted. To compare the performance of $d$th task with other tasks, the normalized root mean square error (NRMSE$_d$) in Eq.~(\ref{eq:NRMSE}) is used, which is given by
\begin{equation}\label{eq:NRMSE}
	{\rm NRMSE}_d = \sqrt{\frac{{\rm MSE}_{d}}{{\rm Var}_{d}}},
\end{equation}
where
\begin{equation}
{\rm Var}_{d} = \frac{1}{k_p-k_t}\sum_{k=k_t + 1}^{k_p} ({y_{d}}(k) - \langle{y_{d}}\rangle)^2
\end{equation}
is the variance in the ground truth ${y_{d}}(k)$ of the task with respect to the temporal mean $\langle{y_{d}}\rangle$ during the test phase $k_t< k<k_p$.

The DPT of L63 comprises of three tasks corresponding to the modes ${A_1, B_1, B_2}$ (refer to Subsection~\ref{rbc}), and the performance of the reservoir topologies for each task of it is shown in Fig.~\ref{fig:mse_modewise}(a). The reservoir state is driven by the input $B_1$ and output matrix is trained to predict either $A_1, B_1, $ or $B_2$. This is a one-step direct prediction of $B_1$ and a cross prediction of $A_1$ or $B_2$. We see from the figure that the direct prediction of $B_1$  performs better with both asymmetric networks. The cross-prediction performance by asymmetric reservoirs is also better for $A_1$. However, the symmetric reservoir topology gives better results for the cross prediction of $B_2$, and the latter is the degree of freedom, that dominates the MSE of the DPT. Thus, we conclude that symmetric reservoirs display a better cross-prediction performance. 

The DPT of L8 consists of eight individual tasks of predicting the modes ${A_1, B_1, B_2, A_2, A_3, B_3, A_4, B_4}$ in Eqs.~\eqref{eq:L8}, out of which modes $A_1$ and $B_1$ are fed as input to the RC model. Figure~\ref{fig:mse_modewise}(b) shows that the direct prediction of $A_1$ and $B_1$ is performed better with the asymmetric (RA) reservoir topology, and the cross prediction of $A_3, B_3, A_4$ and $B_4$ is better accomplished by the symmetric reservoir topology. The overall performance of L8, as shown in Fig.~\ref{fig:mse_comparison}(c) and (f), is thus determined by the magnitude of the NRMSE coming from exactly these cross-prediction tasks. The cross prediction of the remaining modes $A_2$ and $B_2$ is better performed by symmetric (RSS, WSS) and asymmetric (RA) reservoir topologies, respectively.

The differences in performance for the five reservoir topologies are less pronounced for the DPT of SF, which comprises of 9 individual tasks, see  Fig.~\ref{fig:mse_modewise}(c), with respect to both direct- and cross-prediction tasks. This holds in particular for the largest MSE amplitudes. It can be seen that the dominant cross prediction of $a_8$ in this DPT is practically insensitive to the reservoir topology. The DPT of the relatively complex SF system with a larger Kaplan-Yorke dimension $D_{\rm KY}$ shows a slightly better performance with symmetric reservoir topology, see again Fig.~\ref{fig:mse_comparison}(f), which we interpret as the cumulative result of the individual prediction performances of modes in Fig.~\ref{fig:mse_modewise}(c).

\begin{figure}
	\includegraphics[scale=0.75]{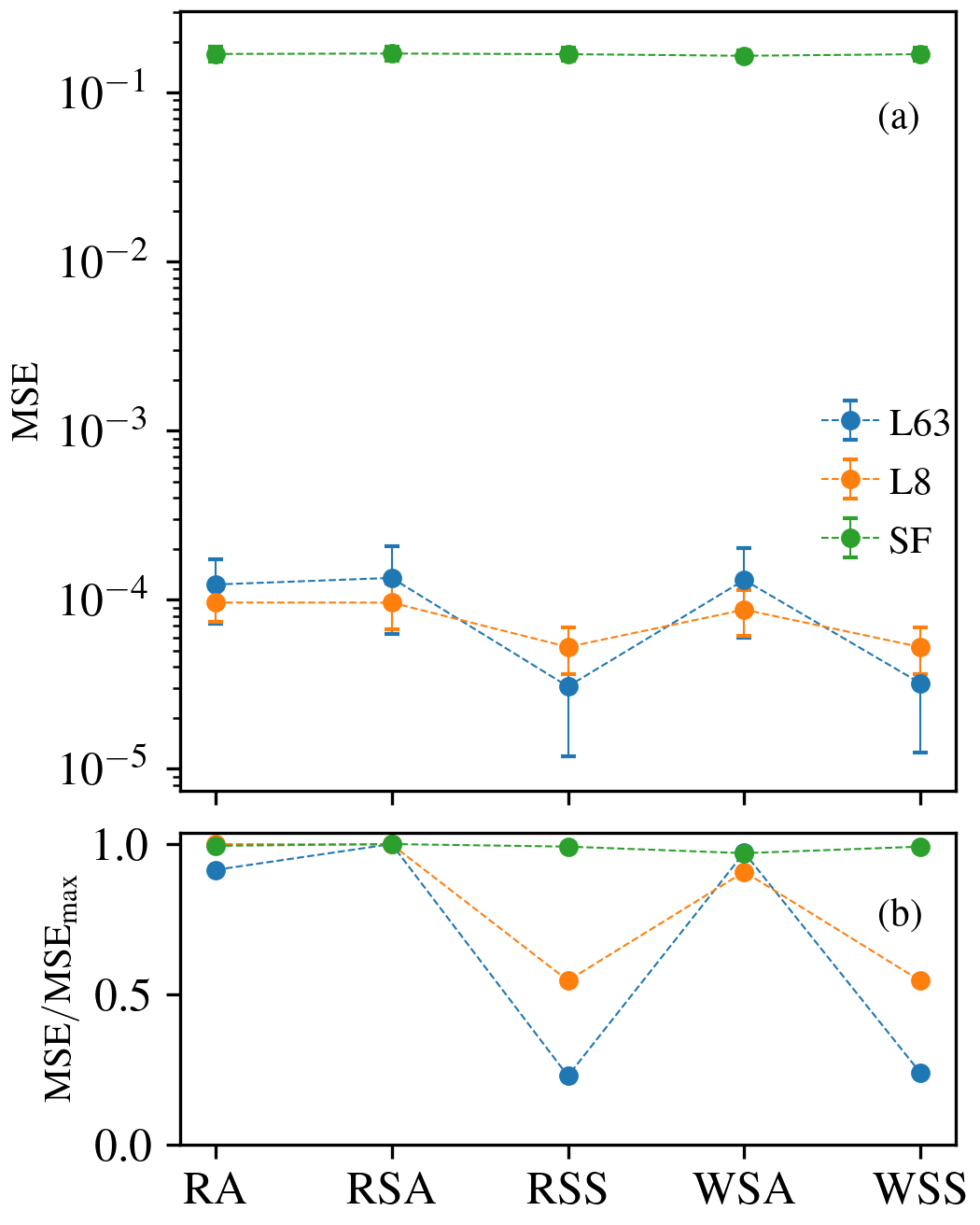}
	\caption{(a) Comparison of MSE for the network performance of the DPT with suboptimal time steps $\Delta t = 0.01 $ for L63, with $\Delta t = 0.02 $ for L8, and $\Delta t = 2.0 $ for SF. All reservoirs have $N = 1024$ nodes. (b) MSE normalized by maximum MSE of each task. Each case for each application, each topology, and each number of reservoirs is obtained as an ensemble median of 1000 different random reservoir network initializations. The legend in (a) applies to both panels.}
	\label{fig:mse_diff_Dt}
\end{figure}

\begin{figure*}
	\includegraphics[scale=0.75]{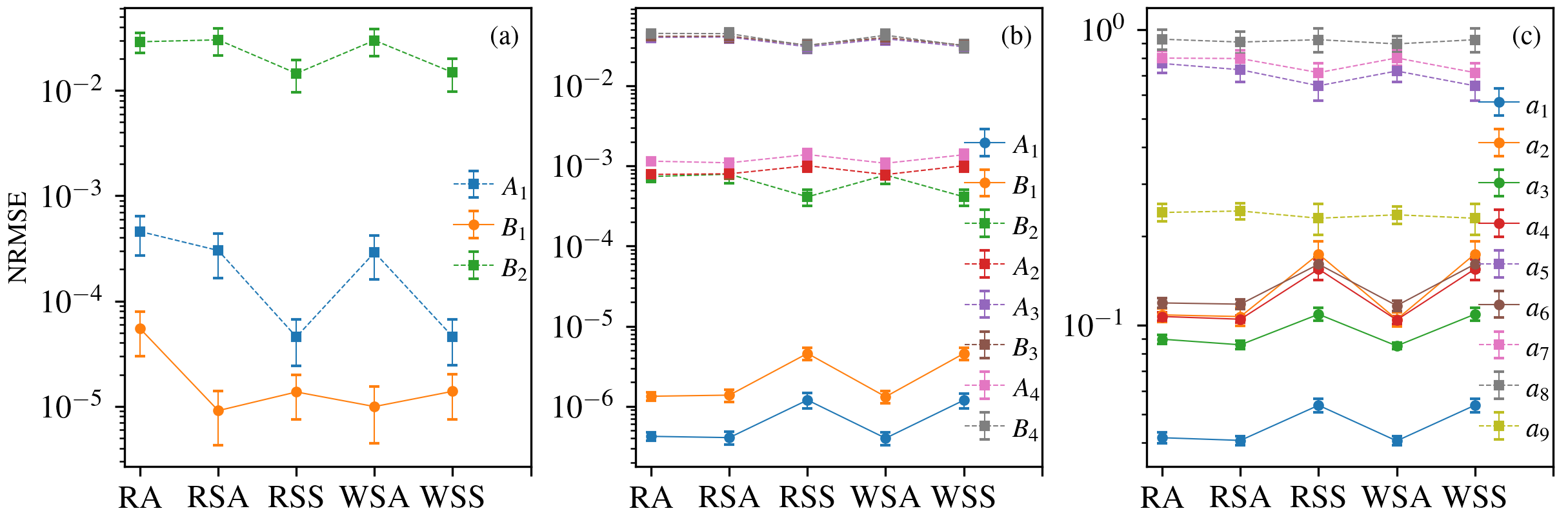}
	\caption{Task-wise comparison of NRMSE for the network performance of the DPT of (a) L63 with $\Delta t = 0.01 $, (b) L8 with $\Delta t = 0.02 $, and (c) SF with $\Delta t = 2.0 $ for $N = 1024$ nodes. Solid lines with filled circle marker represent direct prediction and dashed lines with filled square marker show cross prediction.}
	\label{fig:mse_modewise_diff_Dt}
\end{figure*}

Further, the performance of the networks is investigated for the DPTs with a suboptimal set of reservoir computing time steps $\Delta t = 0.01, 0.02, 2.0$ for dynamical systems L63, L8, and SF, respectively. The results are shown in Fig.~\ref{fig:mse_diff_Dt}. The symmetry of the reservoir network topology is statistically less significant with suboptimal reservoir computing time step, cf. Fig.~\ref{fig:mse_comparison}(f). The trends in individual tasks of the DPTs, however, are similar to those with optimal reservoir computing time step. The direct predictions perform, in general, better with an asymmetric network topology---either RA or RSA and WSA---for all the three systems, as shown in Fig.~\ref{fig:mse_modewise_diff_Dt}. The cross predictions that dominate the MSEs in Fig.~\ref{fig:mse_diff_Dt} tend to be better processed by RC models with symmetric network topology.

%----------------------------------------------------
%----------------------------------------------------
%----------------------------------------------------

\begin{figure}
	\includegraphics[scale=0.75]{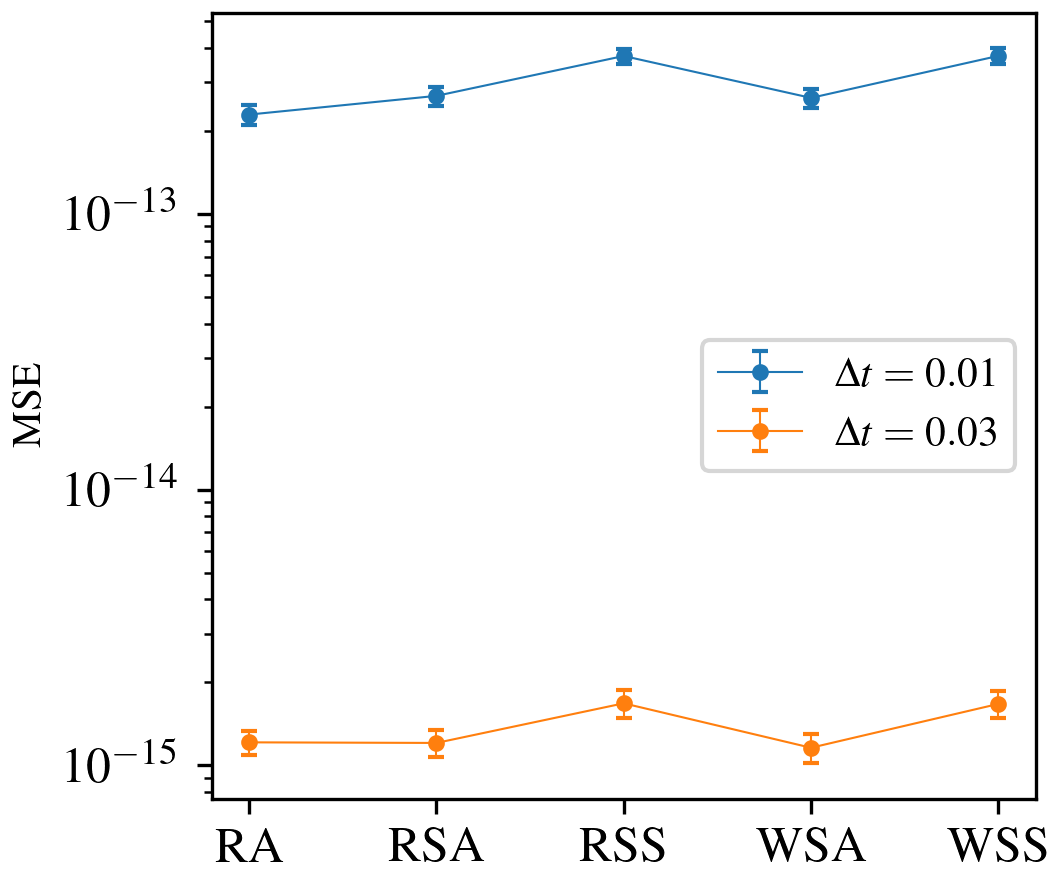}
	\caption{Comparison of MSE for the network performance of the DPT of L63 with $\Delta t = 0.01 $,  and $\Delta t = 0.03 $ for $N = 1024$ nodes. Here, we apply a direct prediction of all 3 degrees of freedom.}
	\label{fig:mse_L63_input_modes_all}
\end{figure}

Finally, we examine the performance and topology relation when full state information is provided as the input to the RC model, i.e., in the absence of cross prediction. We consider the dynamics prediction task of the L63 system only. The RC model receives all three time series $[A_1, B_1, B_2]$ as the input and the optimum reservoir computing time step $\Delta t = 0.03$. The asymmetric random networks learn the dynamics better than their symmetric counterparts irrespective of reservoir computing time step, as shown in Fig.~\ref{fig:mse_L63_input_modes_all}. This was also observed in the prediction of the dynamics described by the Mackey-Glass equation.~\cite{Rathor2025} 
%----------------------------------------------------
%----------------------------------------------------

To summarize our study, the RC model with the optimal sets of hyperparameters in the open-loop scenario perform differently well for the five network topologies. In Fig.~\ref{fig:mse_comparison}, the RA topology, which we identified in Ref.~\onlinecite{Rathor2025} as the best performing network, is typically not the best reservoir topology for tasks with more degrees of freedom, which involve cross-prediction tasks. The statement, that the asymmetric reservoir learns best, is thus not generalizable and depends on the specifics of the dynamical prediction task. For the tasks with $N_{in}<N_{out}$, due to cross prediction, we find that symmetric networks perform generally better.
\section{Conclusions and Outlook}
The first main goal of the present study was to shed new light on the connection between reservoir computing model performance and topology of the central building block of this class of recurrent machine learning algorithms, the random reservoir network. To this end, we set up a systematic construction scheme that can {\em separately} prescribe, both the symmetry of network connectivity and weights along the active node connections. Secondly, we wanted to carry this study to applications beyond the almost exclusively used academic benchmarks of Mackey-Glass time-delayed and Lorenz 63 model dynamics. It actually turned out, that the two application cases with more degrees of freedom, an extended Lorenz-type model and a three-dimensional shear flow,  showed a decreasing susceptibility of the specific reservoir network topology to the learning performance. 

More detailed, three topologies of networks in RC, that are conceived on the basis of the node connections and their weights, were investigated. These topologies are R-A, RS-A, and RS-S, which are constructed from randomly initialized connection and weight matrices. These topologies, along with two more random network types constructed from the Watts-Strogatz framework with a rewiring probability of $p=1$, were taken. They were termed  WS-A and WS-S. The latter two topologies are similar to RS-A and RS-S, except for a difference in the width of the node-degree distribution.\cite{Rathor2025} These topologies are used to predict the dynamics of three fluid flow systems, all of which can be described by a Galerkin-type nonlinear dynamical systems model. These reduced-order models prescribe the spatial modes in correspondence with the boundary conditions of the fluid flow problem and describe the dynamics of the expansion coefficients, which get coupled in a system of nonlinear ordinary differential equations. The tasks differ by increasing order of dimensionality (or number of degrees of freedom) and complexity, the latter of which is quantified by the Kaplan--Yorke dimension. These systems are L63, L8, and SF with dimensions 3, 8, and 9 respectively. A one-dimensional time series representing the dynamics from a Mackey--Glass equation at a fixed delay time $\tau$ is additionally considered for comparison with the aforementioned systems.

Our open-loop one-step prediction scenario is as follows: the RC predicts the dynamics, which comprises all degrees of freedom of the target nonlinear dynamical system, with partial information at the input to the RC model. We find a generic trend, namely, that the dynamics prediction is performed better if symmetric random network topology (RS-S or WS-S) is used. This holds for the three models that describe two-dimensional Rayleigh-B\'{e}nard convection (L63, L8) and three-dimensional shear flow (SF). A disentanglement of the prediction into individual direct- and cross-prediction tasks revealed the following results: (i) the contribution to the total MSE coming from the direct-prediction pipelines is negligible in comparison to those coming from cross predictions; (ii) cross-prediction subtasks thus determine the overall performance. They are the ones, that require a delay-embedding, i.e. a looping of the information inside the network which is a short-term memory. This finding is complemented by several previous works, which have shown that performance can be improved using pre- or post-processing which introduces explicit delay.~\cite{Marquez_2019,Jaurigue_2025,Jaurigue2024,Fleddermann_2025}  Here, we observed that exactly these cross-prediction subtasks are better carried out by symmetric networks, RS-S or WS-S, even though the performance decreases with increasing learning complexity, which we quantify by $D_{\rm KY}$. Thus we conclude from the current study, that symmetric networks should be chosen for the dynamics prediction of nonlinear dynamical systems with larger numbers of degrees of freedom and $N_{\rm in}<N_{\rm out}$.

The observed trend that the network becomes increasingly insensitive to its topology as learning complexity grows, raises new questions for spatio-temporal pattern recognition tasks. Our findings suggest that the current hard-wired, static RC random network architecture is overly simplistic, and that incorporating plasticity into the RC network would likely be necessary to improve DPT performance.

\section*{Acknowledgments} 
The work of S.K.R. and J.S. is funded by the European Union (ERC, MesoComp, 101052786). Views and opinions expressed are however those of the authors only and do not necessarily reflect those of the European Union or the European Research Council. M.Z. acknowledges financial support by Deutsche Forschungsgemeinschaft (DFG, German Research Foundation) – Project RECOMMEND Project number 536063366 as part of the DFG priority program DFG-SPP 2262 MemrisTec – Project number 422738993. L.J. is funded by the Carl-Zeiss-Stiftung.

\section*{AUTHOR DECLARATIONS}
\textbf{Conflict of Interest} 
The authors have no conflicts to disclose.

\bibliography{main}

% ========================= Main text ends here =========================

\appendix
\section{Further results for SF case}
In appendices A and B, we present more results of the DPTs for the two dynamical systems with the highest number of degrees of freedom, SF in this appendix section, and L8 in appendix B. While with increasing number of degrees of freedom the RC will not be able anymore to predict the time series for long times, it can reproduce statistical correlations, as we will demonstrate in the following. To this end, we reconstruct from the time series $A_1(t)$ to $B_4(t)$ in L8 and $a_1(t)$ to $a_9(t)$ in SF the time-dependent velocity and temperature fields to analyse their statistics similar to what was done in refs.\cite{Srinivasan2019,Heyder2021,Heyder2022}
%-----------------------------------------------
\begin{figure*}[b!]
	\includegraphics[scale=0.8]{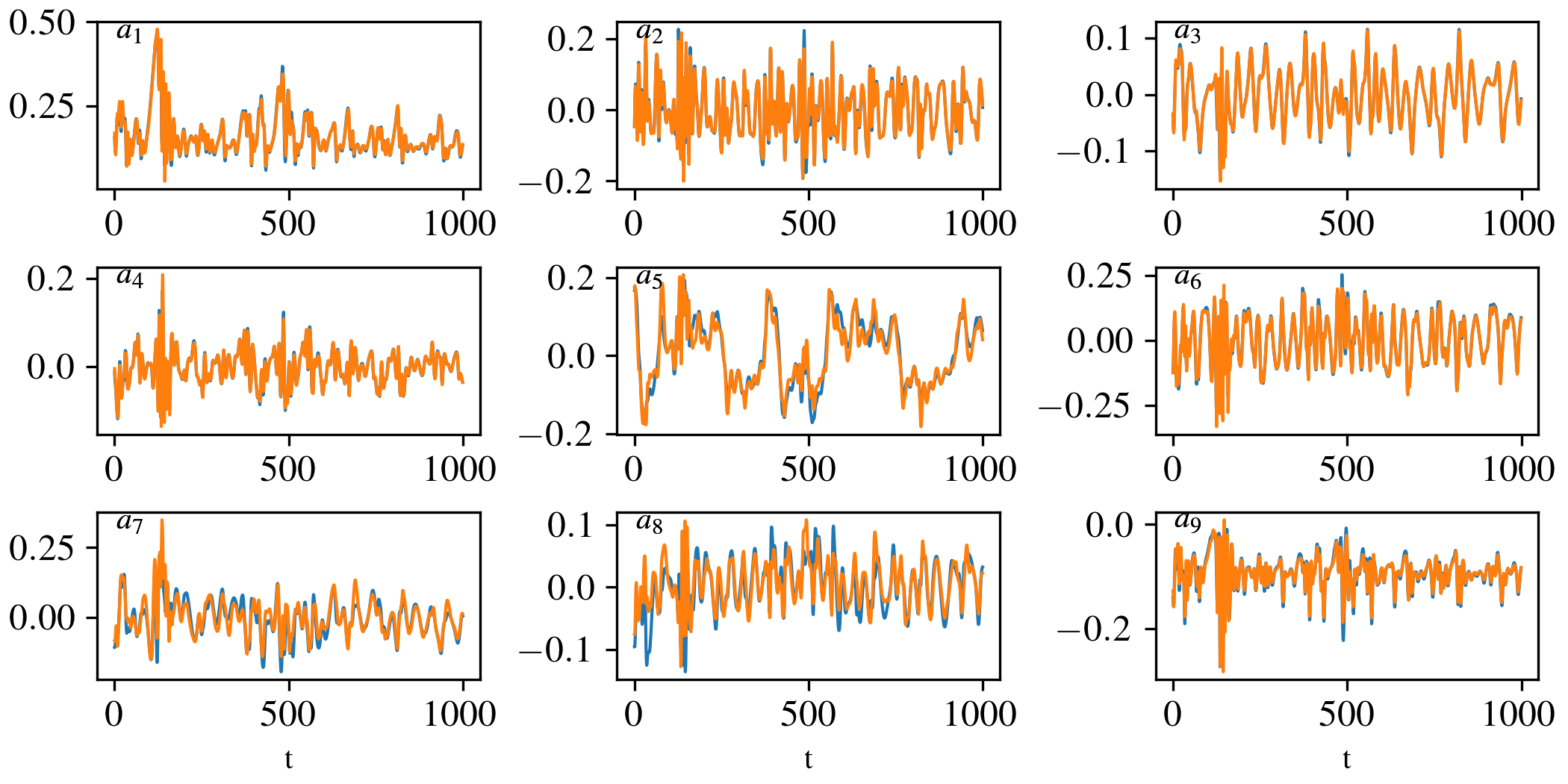}
	\caption{Predicted time series, $a_1(t)$ to $a_9(t)$ of the three-dimensional plane shear flow (SF) model in Eqs. \eqref{eq:L9} with a reservoir computing model utilizing the R-A network topology. Solid blue line represents the ground truth, and solid orange line shows the prediction. The number of input modes is five in open-loop prediction mode. These modes are $a_1, a_2, a_3, a_4$ and $a_6$. Reconstruction by the reservoir is for all nine degrees of freedom.}
	\label{fig:prediction_SF}
\end{figure*}
%-----------------------------------------------
\begin{figure*}
	\includegraphics[scale=0.9]{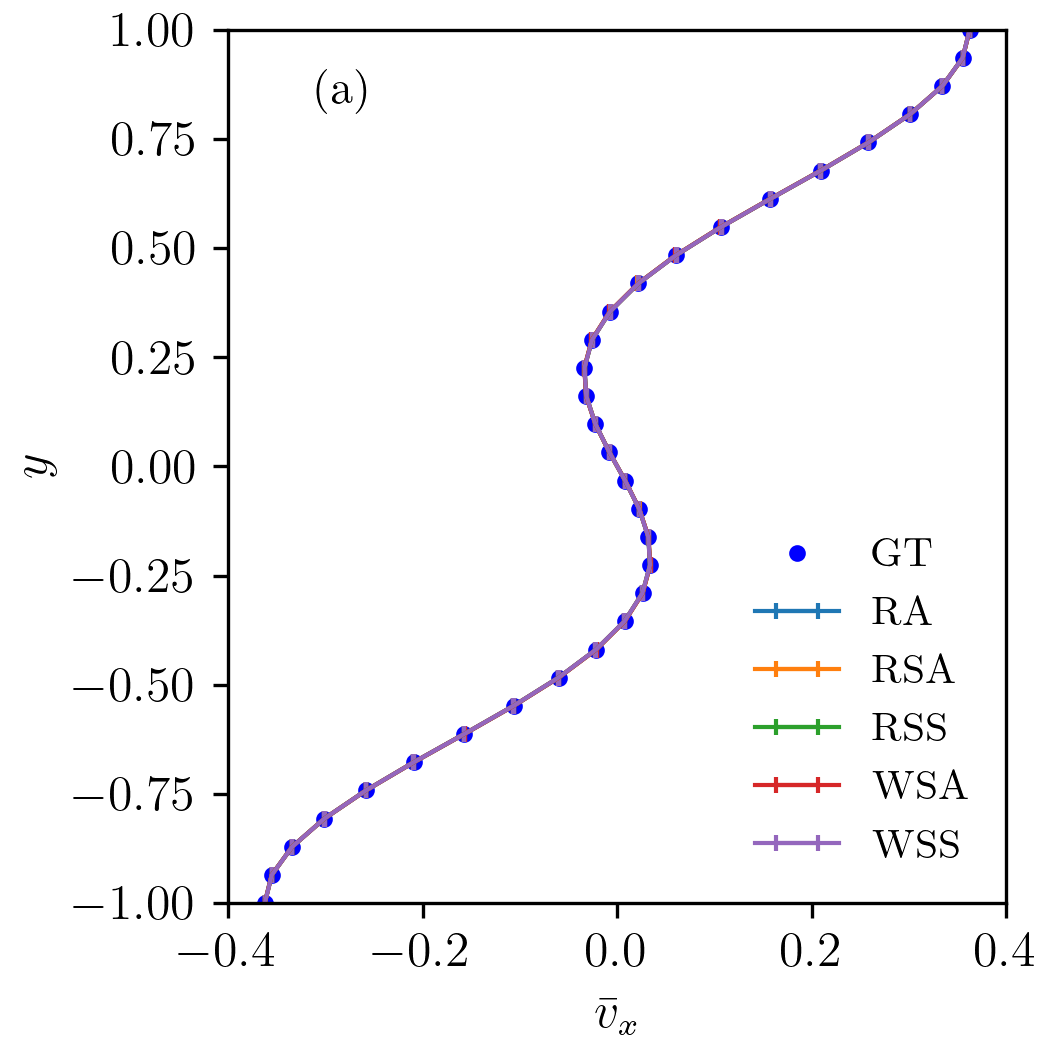}
	\includegraphics[scale=0.9]{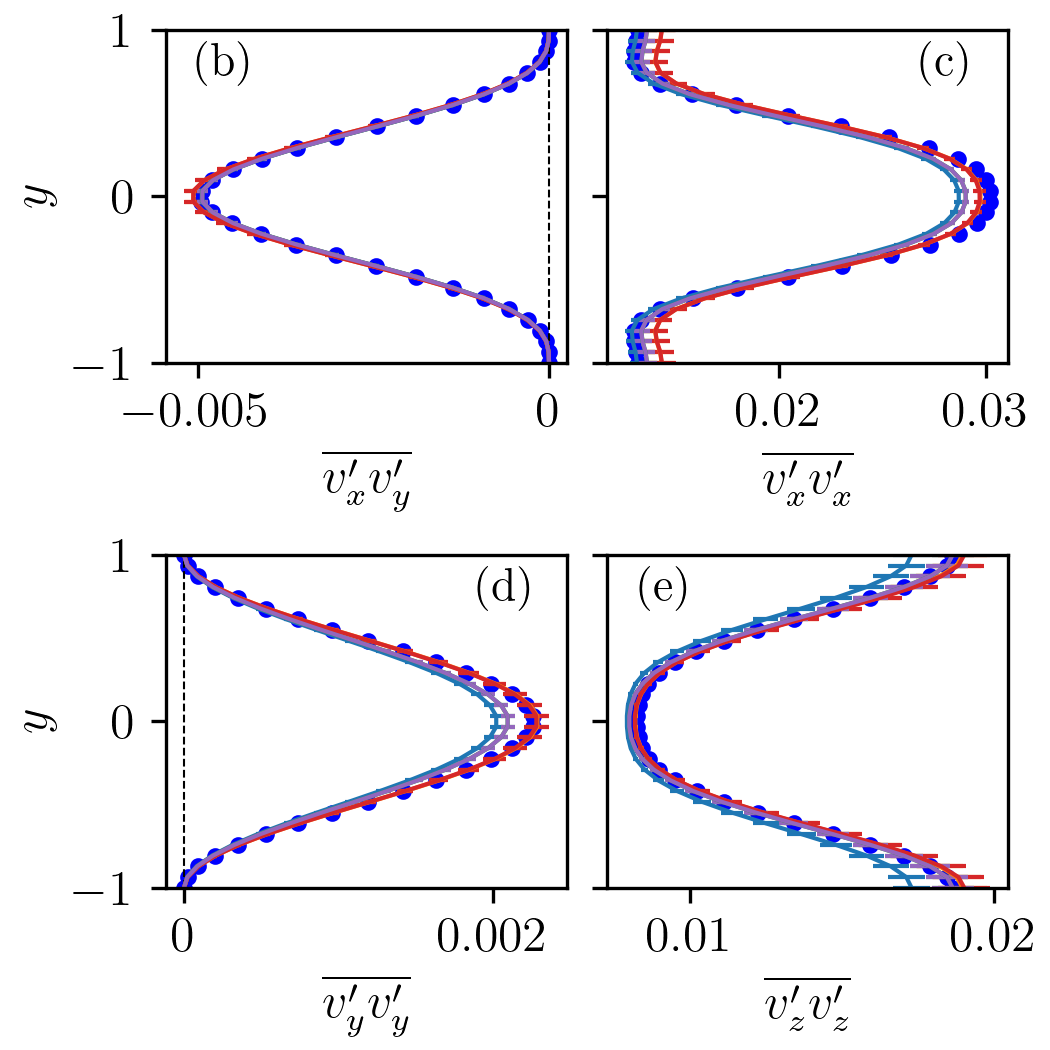}
	\caption{Fluid flow statistics of the three-dimensional shear flow model. (a) Comparison of predicted mean streamwise velocity component proﬁle obtained from different networks with the ground truth (GT). (b,c,d,e) Comparison of predicted mean Reynolds stress proﬁles for different networks with the GT. All profiles are averages with respect to time as well as streamwise and spanwise directions. The error bars are the standard deviations over predictions from 100 differently realized reservoir matrices for a given topology. Note that the Reynolds stress profiles $\overline{v'_yv'_y}$ and $\overline{v'_xv'_y}$ have to be zero at $y=\pm 1$ due to the free-slip boundary conditions for the velocity field.}
	\label{fig:mean_flow}
\end{figure*}
%-----------------------------------------------

Figure~\ref{fig:prediction_SF} displays the performance of the R-A network for this three-dimensional shear flow model. It can be observed, how the prediction starts to deviate from the ground truth for the degrees of freedom that have to be reconstructed, caused by the sensitivity to small deviations by round-off errors. This example already demonstrates that we cannot expect to predict a specific trajectory, as being the case in the L63 system.\cite{Pfeffer2022} 

We probe the accuracy of the prediction of the statistical properties of SF, such as mean basic velocity profile, $\bar{v}_x(y)$, in the wall-normal direction. Such quantities are in view of applications more important than individual system trajectories. Time-dependent flow statistics can be quantified by the following normalized relative error (NARE) \cite{Srinivasan2019,Pandey2020}, which is defined as  
\begin{equation}
	E[q(y)] = \frac{1}{2 \textrm{max}_{y \in [-1,1]}} \int_{-1}^{1} \vert q^{GT}(y) - q^{P}(y)\vert \textrm{d}y\,.
	\label{eq:NARE_q}
\end{equation}
Here, $q$ is the quantity of interest, and $q^{GT}$ and $q^p$ are the ground truth and prediction. Table~\ref{tab:NARE_summary} shows that the outcome with respect to NARE is slightly different to that for the MSE. With respect to a reconstruction of the shear flow in a statistical sense, the R-A network performs best, with $E[\bar{v}_x(y)] = 0.05\%$.

%-----------------------------------------------
\begin{table*}
	\begin{tabularx}{\textwidth}{ 
			>{\hsize=.5\hsize\raggedright\arraybackslash}X 
			>{\centering\arraybackslash}X 
			>{\centering\arraybackslash}X
			>{\centering\arraybackslash}X
			>{\centering\arraybackslash}X
			>{\hsize=.5\hsize\centering\arraybackslash}X
		}
		\hline\hline
		& $E[\bar{v}_x(y)]$ & $ E[\overline{v_x' v_y'}(y)]$ & $ E[\overline{v_x' v_x'}(y)]$ & $ E[\overline{v_y' v_y'}(y)]$& $ E[\overline{v_z' v_z'}(y)]$\\
		\hline
		RA   & $0.05 \pm 0.03$ & $0.5 \pm	 0.3$ & $3.0 \pm	 0.3$ & $5.1 \pm 0.7$ & $4.5 \pm 1.4$ \\
		RSA  & $ 0.08 \pm 0.05$ & $ 0.9 \pm 0.5 $ & $ 2.0 \pm 0.6 $ & $ 1.1 \pm 0.7 $ & $ 1.6 \pm 0.9 $\\
		RSS  & $ 0.06 \pm 0.03 $& $ 0.4 \pm 0.3 $ & $ 2.4 \pm 0.2 $ & $ 3.5 \pm 0.7 $ & $ 1.6 \pm 0.7 $\\
		WSA  & $ 0.07 \pm 0.04 $ & $ 1.1 \pm 0.6$ & $ 2.0 \pm 0.5 $ & $ 1.2 \pm 0.6 $ & $ 2.0 \pm 0.9 $\\
		WSS  & $ 0.06 \pm 0.03$ & $ 0.4 \pm 0.3 $ & $ 2.4 \pm 0.2 $ & $ 3.5 \pm 0.7 $ & $ 1.6 \pm 0.7 $\\
		\hline\hline
	\end{tabularx}
	\caption{Median of normalized relative error (NARE), as an additional measure of performance next to the MSE. The error is calculated as the median absolute deviation (MAD) of predictions from 100 differently realized reservoir matrices for a given topology. The quantity is defined by Eq.~\eqref{eq:NARE_q} for the Galerkin model of a three-dimensional shear flow (SF), which is run in an open loop scenario. Five different quantities are investigated, the streamwise mean flow and 4 important components of the Reynolds stress tensor. See also the corresponding mean vertical profiles, which are displayed in Fig.~\ref{fig:mean_flow}.}
	\label{tab:NARE_summary}
\end{table*}
%-----------------------------------------------
\begin{figure*}
	\includegraphics[scale=0.9]{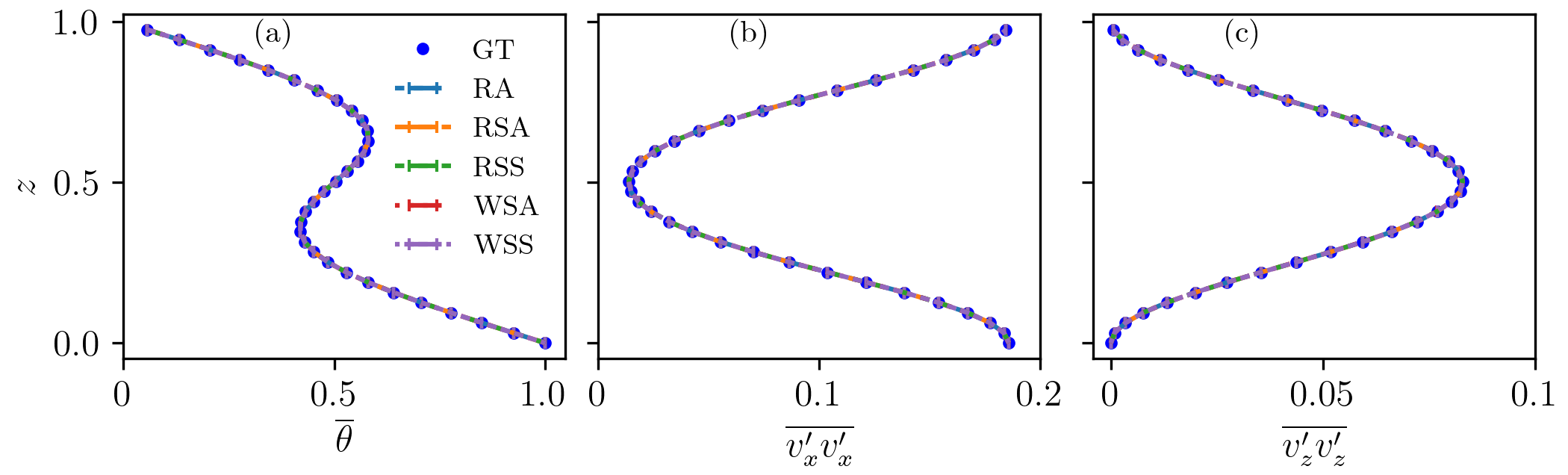}
	\caption{Statistics of the extended Lorenz-type model in Eqs.~\eqref{eq:L8}. (a) Comparison of predicted mean thermal fluctuations obtained from different networks with the ground truth (GT). (b,c) Comparison of predicted mean Reynolds stress proﬁles for different networks with the GT. All profiles are averages with respect to time as well as $x-$direction. The error bars are the standard deviations over predictions from 100 differently realized reservoir matrices for a given topology. The Reynolds stress profiles $\overline{v'_zv'_z}$ have to be zero at $z=0, 1$ due to the free-slip boundary conditions.}
	\label{fig:mean_flow_of_L8}
\end{figure*}

Figure~\ref{fig:mean_flow} displays some important statistical correlations of SF. The vertical {(wall-normal) mean flow profile in panel (a) is well reconstructed for all five cases. In panels (b--e) of the same figure, we provide the vertical mean profiles of four components of the Reynolds stress tensor $\overline{v_i' v_j'}$  in Eq.~(\ref{eq:RS}), which quantifies the correlations between the velocity components. It is seen that first- and second-order moments of the velocity fields are reproduced fairly well by the RC for all five topologies. The strongest scatter arises for the autocorrelations of the spanwise velocity component with the prediction by R-A topology closest to the ground truth. The components of the Reynolds stress tensor are given by 
	\begin{equation}
		\overline{v_i' v_j'}(y) = \frac{1}{L_x L_z}\int_{0}^{L_z}\int_{0}^{L_x}\langle v_i' v_j'\rangle_t \textrm{d}x \textrm{d}z\,.
		\label{eq:RS}
	\end{equation}
They are obtained by the Reynolds decomposition of the velocity field components 
	\begin{equation}
		v_i'(x,y,z,t) = v_i(x,y,z,t) - \langle v_i(x,y,z)\rangle_t\,, 
	\end{equation}
where $\langle \cdot\rangle_t$ denotes averaging with respect to time.

\section{Further results for L8 case}
For completeness, we repeated this analysis for the L8 model. Figure~\ref{fig:mean_flow_of_L8} reports some results for the L8 model. We show the mean temperature profile taken across the convection layer together with the profiles of the velocity fluctuations in $x$ and $z$ directions, the diagonal entries of the Reynolds stress tensor. A very good agreement of all 5 networks with the ground truth can be observed. The number of nodes in all reservoir networks was $N=1024$. The standard deviation error from the predictions of 100 differently initialized reservoir coupling matrices is $O(10^{-6})$. We see in both cases that the RC models can be used as surrogate models for the fluid flows.

\end{document}